\documentclass[times,twocolumn,final,longtitle]{elsarticle}

\usepackage{jasr}
\usepackage{framed,multirow}

\usepackage{amssymb}
\usepackage{latexsym}

\usepackage{url}
\usepackage{xcolor}
\definecolor{newcolor}{rgb}{.8,.349,.1}

\usepackage[citebordercolor=white,colorlinks=true,citecolor=blue]{hyperref}

\usepackage{romannum}

\newcommand{\ionn}[2]{\textup{#1\,\rmfamily\scriptsize{\Romannum{#2}}}}
\newcommand\farcs{\mbox{$.\!\!^{\prime\prime}$}}%

\newcommand\arcsec{\ensuremath{^{\prime\prime}}}

\begin{document}
\pagenumbering{arabic}

\verso{Louis}

\begin{frontmatter}


\title{Classification of circular polarization Stokes profiles in a sunspot using 
\textit{k-means} clustering}%

\author[1]{Rohan Eugene \snm{Louis}\corref{cor1}}
\cortext[cor1]{Corresponding author: 
  Tel.: +91-294-2457-217;  }
\ead{rlouis@prl.res.in}
\author[1]{Shibu K. \snm{Mathew}}  
\author[1]{A. Raja \snm{Bayanna}}

\address[1]{Udaipur Solar Observatory, Physical Research Laboratory, Dewali Badi Road, Udaipur - 313001, Rajasthan, India}

\received{26 July 2023}
\finalform{20 Dec 2023}
\accepted{20 Dec 2023}
\availableonline{--}
\communicated{--}

\begin{abstract}
The magnetic and velocity fields in sunspots are highly structured on small spatial scales which are 
encoded in the Stokes profiles. The Stokes profiles are in turn, derived from a sequence of polarization 
modulations on the incoming light that are imaged using an analyser-detector combination. Our aim is to identify 
Stokes profiles in a sunspot which exhibit spectral characteristics that deviate from those associated with the 
Evershed flow and their corresponding spatial distribution. To that end, we employ the 
{\tt{k-means}} clustering routine to classify Stokes $V$ spectra in the penumbra of a regular, unipolar sunspot, 
that also comprises a granular and a filamentary light bridge. 
We find that 75\% of the penumbral region, corresponding to about 93500 pixels, is dominated by profiles comprising 
two, nearly anti-symmetric lobes, while 21\% of the area is occupied by three-lobed profiles that are associated 
with the Evershed flow returning to the solar photosphere. 
The remaining 4\% of the penumbral area is dominated by four groups/families of profiles - 
Group 1: three-lobed profiles in which both the rest and strong downflowing (sometimes supersonic)
component have the same polarity as the host sunspot and seen exclusively in the filamentary light bridge.
Group 2: single, red-lobed profiles occupying 
an area of about 2\% located at the outer penumbra in discrete patches that possibly signify the 
downflowing leg of an $\Omega$-loop.  Group 3: three-lobed
or highly asymmetric profiles, in which the rest component and the strong downflowing component 
have an opposite polarity as the sunspot. These occupy about 1.4\% of the penumbra's area and are seen in conspicuous, 
elongated structures or isolated patches in the outer penumbra and at the penumbra-quiet Sun boundary. Group 4: 
three lobed-profiles, in which the rest component has the same polarity as the sunspot and a weaker, upflowing 
component with a polarity opposite that of the sunspot. These profiles are located close to the entrance of the 
filamentary light bridge and are found in only 0.12\% of the penumbral area. These minority groups of profiles 
could be related to dynamic phenomena that could also affect the overlying chromosphere. The 
simplicity and speed of {\tt{k-means}} can be utilized to identify such anomalous profiles in larger data sets 
to ascertain their temporal evolution and the physical processes responsible for these inhomogeneities.
\end{abstract}


\end{frontmatter}



\section{Introduction} 
\label{intro}
Sunspots have been extensively studied for more than four centuries firmly
confirming their intricate connection to the solar activity cycle.
The subsequent discovery of the Zeeman effect \citep{1897Natur..55..347Z} 
aided in establishing the presence of magnetic fields in sunspots 
\citep{1908ApJ....28..315H,1919ApJ....49..153H} that heralded a paradigm 
shift in solar physics. Over the last 40 years, advancements in ground- and 
space-based spectro-polarimetric observations have enabled us to infer magnetic and 
velocity fields in sunspots with very high spatial and temporal resolution. 

The inference of the photospheric magnetic field is done from the various polarization
states of light which can be described in terms of the four Stokes parameters 
\citep[\textit{I}, \textit{Q}, \textit{U}, \textit{V};][]{1852RSPT..142..463S}
and solving the radiative transfer equation under the assumption of local thermodynamic
equilibrium.  
Stokes \textit{Q},
and \textit{U} describe the difference of linearly polarized light at 0$^\circ$ and 90$^\circ$, 
and 45$^\circ$ and 135$^\circ$, respectively. On the other hand, Stokes \textit{V} is the 
difference of the right- and left-circular polarization states of light. 
Stokes \textit{I} represents the total intensity of the incoming radiation.
The four Stokes 
parameters can thus provide information of the thermal, magnetic, and kinematic conditions in 
the solar atmosphere where the observed spectral lines form. The extraction of the
physical parameters from the Stokes profiles is currently done using spectral inversion codes 
\citep[ICs;][]{2001ASPC..236..487S,2006ASPC..358..107B,delToroIniesta16}, 
which vary in complexity and approximations of the solar atmosphere. 
The simplest and fastest
of ICs is based on the Milne-Eddington (ME) approximation which assumes that all physical 
parameters remain constant with height and the source function varies linearly with optical depth
\citep{1956PASJ....8..108U,1962IzKry..28..259R,1972SoPh...27..319L,1987ApJ...322..473S}.
Other ME-based ICs include the
MILne–Eddington inversion of the pOlarized Spectra 
\citep[MILOS;][]{2007ApJ...662L..31O,2007ApJ...670L..61O},
MERLIN \citep{2007MmSAI..78..148L},  
Very Fast Inversion of the Stokes Vector \citep[VFISV;][]{2011SoPh..273..267B}, and 
Stokes Profile INversion \cite[SPIN;][]{2017SoPh..292..105Y}. However, ME-based ICs
cannot handle gradients along the line-of-sight that produce 
asymmetric Stokes profiles. Other ICs which can account for gradients along the line-of-sight
as well as multiple components within the resolution element, include the
Stokes inversion based on Response Functions \citep[SIR;][]{1992ApJ...398..375R}, 
MIcro Structured Magnetic Atmospheres \citep[MISMA;][]{1997ApJ...491..993S},   
SIRGAUSS \citep{2003ASPC..307..301B,2007PASJ...59S.601J}, and SIRJUMP \citep[employed in][]{2009ApJ...704L..29L}.
The main characteristics of the ICs mentioned above is that they consider local thermodynamic 
equilibrium (LTE), which is a valid approximation in the solar photosphere.
They have been widely utilized to retrieve the atmospheric conditions in the solar
photosphere where the Stokes profiles are highly asymmetric or anomalous as we shall see below.

In the absence of gradients in the magnetic field and line-of-sight (LOS) velocity, the 
Stokes \textit{V} profile comprises two, anti-symmetric, equi-amplitude lobes which 
are generally seen in the sunspot umbra and a large part of the penumbra. However, 
the difference between the blue and red lobe amplitudes, or amplitude asymmetry, is 
typically non-zero and exhibits a sign change along the radial penumbral distance, 
wherein the blue lobe is larger (smaller) than the red lobe in the inner (outer) penumbra
\citep{2002A&A...381..668S}.
Furthermore, Stokes \textit{V} profiles with a third, additional lobe having a polarity opposite that of 
the parent spot and strongly-red shifted are seen in many patches located in the mid- and outer-
penumbra. Such profiles are associated with the Evershed flow \citep{1909MNRAS..69..454E} 
returning to the solar photosphere at supersonic speeds and have been observed in the visible 
\citep{1997Natur.389...47W} as well as infra-red \citep{2001ApJ...549L.139D}. 
Such profiles have been analysed using different atmospheric configurations in spectral ICs, 
either using two height-independent magnetic components within the resolution element 
\citep{2004A&A...427..319B,2008A&A...480..825B},
or, gradients in the physical parameters along the line-of-sight
\citep{2003A&A...410..695M,2010ASSP...19..193B,2013A&A...549L...4R}. 
The association of the returning downflows in sunspot penumbrae 
with overturning convection, as predicted by simulations \citep{2009Sci...325..171R},
has been a topic of extensive research over more than two decades 
\citep{1999A&A...349L..37S,2007ApJ...658.1357S,2009A&A...508.1453F,2011Sci...333..316S,
2011ApJ...734L..18J,2013A&A...549L...4R,2013A&A...550A..97F,2015ApJ...803...93E,2016A&A...596A...4F}. 
Detailed reviews on this topic are available in 
\citet[][ and references therein]{2009SSRv..144..213S} and 
\citet[][ and references therein]{2011LRSP....8....4B}.

Along with the supersonic penumbral downflows, the source of the Evershed flow can be 
traced to bright penumbral grains at the heads of filaments at the umbra-penumbra
boundary \citep{2006ApJ...646..593R,2013A&A...550A..97F} which are associated with weak blue-shifts
and comprise a weakly asymmetric blue lobe in the Stokes \textit{V} profile
\citep{2007PASJ...59S.593I,2013A&A...550A..97F}. On the other hand, spectral profiles above 
sunspot light bridges (LBs) have been known to be asymmetric 
\citep{1994ApJ...426..404S,2002ApJ...576.1048S,2006A&A...453.1079J}, 
indicating gradients along the LOS, as well as anomalous, wherein
the Stokes \textit{V} profiles comprise three lobes, with the third lobe
having the same polarity as the sunspot  and indicative of supersonic
downflows \citep{2009ApJ...704L..29L}. The latter, in particular,
are also seen at the umbra-penumbra boundary of sunspots \citep{2011ApJ...727...49L}.
The supersonic flows in LBs can also be blue- or red-shifted to near-supersonic speeds 
with a polarity opposite to that of the spot and comprise a small fraction of the resolution 
element \citep{2015AdSpR..56.2305L}.

With the spatial resolution of ground-based telescopes approaching, or becoming smaller than, 
the photon mean free path of about 100\,km in the photosphere \citep{2015SoPh..290..979J}, 
one expects to encounter a variety of inhomogeneities in the physical parameters recovered
from a sunspot. 
In the conventional setup of ICs, the merit function ($\chi^2$) between the observed and synthetic Stokes
profiles is iteratively minimized using the Marquardt algorithm \citep{Press1986}. The iterations 
involve the modification of an initial depth-dependent model atmosphere supplied by the user 
which make up the elements of the propagation matrix in the radiative transfer equation (RTE). 
The elements of the propagation matrix comprise the absorption, emission, and birefringence phenomena 
which characterize the medium \citep{2007insp.book.....D}. The synthetic profiles are then generated 
by numerically integrating the RTE for the wavelength range present in the observations. 
The iterative minimization of the $\chi^2$ is achieved by determining the perturbations of the 
guess model atmosphere such that the synthetic profiles match the observed ones. Under 
a linear approximation, the sensitivities of the Stokes profiles to perturbations of the 
physical parameters are called Response Functions \citep{1992ApJ...398..375R} that are necessary to 
compute the partial derivatives of $\chi^2$ with respect to the model atmospheric parameters. In order
to reduce the number of free parameters, the perturbations of the depth-dependent physical parameters
are determined only at discrete points in optical depth called nodes. The number of nodes can be set 
independently for each physical parameter. Thus, the IC delivers a model atmosphere for the chosen set of 
nodes \citep{Luis2003}. One should note that, the solution obtained by this method is not unique, and 
during the minimization iterative process a local minimum may be reached. 
As we have seen above, the number of nodes, which
also represent the degrees of freedom, have to be specifically tailored for each type
of inhomogeneous atmospheric stratification while also ensuring its physical realism and 
feasibility. 
Thus, an apriori knowledge of the spectral shape would assist in optimizing the set of nodes in an 
IC to adequately determine the atmospheric stratification.  

In this article, we classify and analyze Stokes \textit{V} profiles in a sunspot
based on their shapes, similar to the strategy adopted by \citet{2011A&A...530A..14V} for quiet-Sun (QS) 
profiles. The aim of this study is to determine the fraction of $V$ profiles that deviate from 
the set of well known profiles in a sunspot, the number of families/groups of such profiles, their 
preferential spatial distribution, if any, and the atmospheric stratification associated with them. 
The rest of the paper is organized as follows, Sections~\ref{obs} and \ref{analysis} describes 
the observations used and the data analysis respectively. We present our results in 
Section~\ref{results} with Sections~\ref{discuss} and \ref{conclude} reserved for Discussion and 
Conclusions, respectively.

\section{Observations}
\label{obs}
The sunspot under study was the leading spot in NOAA active region (AR) 10953
that was observed by Hinode \citep{2007SoPh..243....3K} on 2007 May 01 when the 
AR was at a heliocentric angle of 8$^\circ$. We use the  
spectropolarimeter \citep[SP;][]{2001ASPC..236...33L,2008SoPh..249..233I} 
of the 50-cm Solar Optical Telescope \citep[SOT;][]{2008SoPh..249..167T}
on board Hinode. After processing the data obtained on board, Hinode
provides the four Stokes profiles of the \ionn{Fe}{1} lines at 630\,nm 
with a spectral sampling of 21.55\,m\AA, a step width of 0\farcs15, a pixel size 
of 0\farcs16 along the slit, and an exposure time of 4.8\,s per slit position 
using the normal map mode from 10:46--12:25\,UT. 
Out of the total SP field-of-view (FOV) of 150\arcsec$\times$164\arcsec\, we selected
a FOV of 67.5\arcsec$\times$72\arcsec\, around the sunspot.
The observations were corrected for dark current, 
flat field, thermal flexures, and instrumental polarization using routines 
included in the Solar-Soft package \citep{2013SoPh..283..601L} to yield Level-1 data.

\begin{figure*}[!ht]
\centerline{
\includegraphics[angle=90,width = \textwidth]{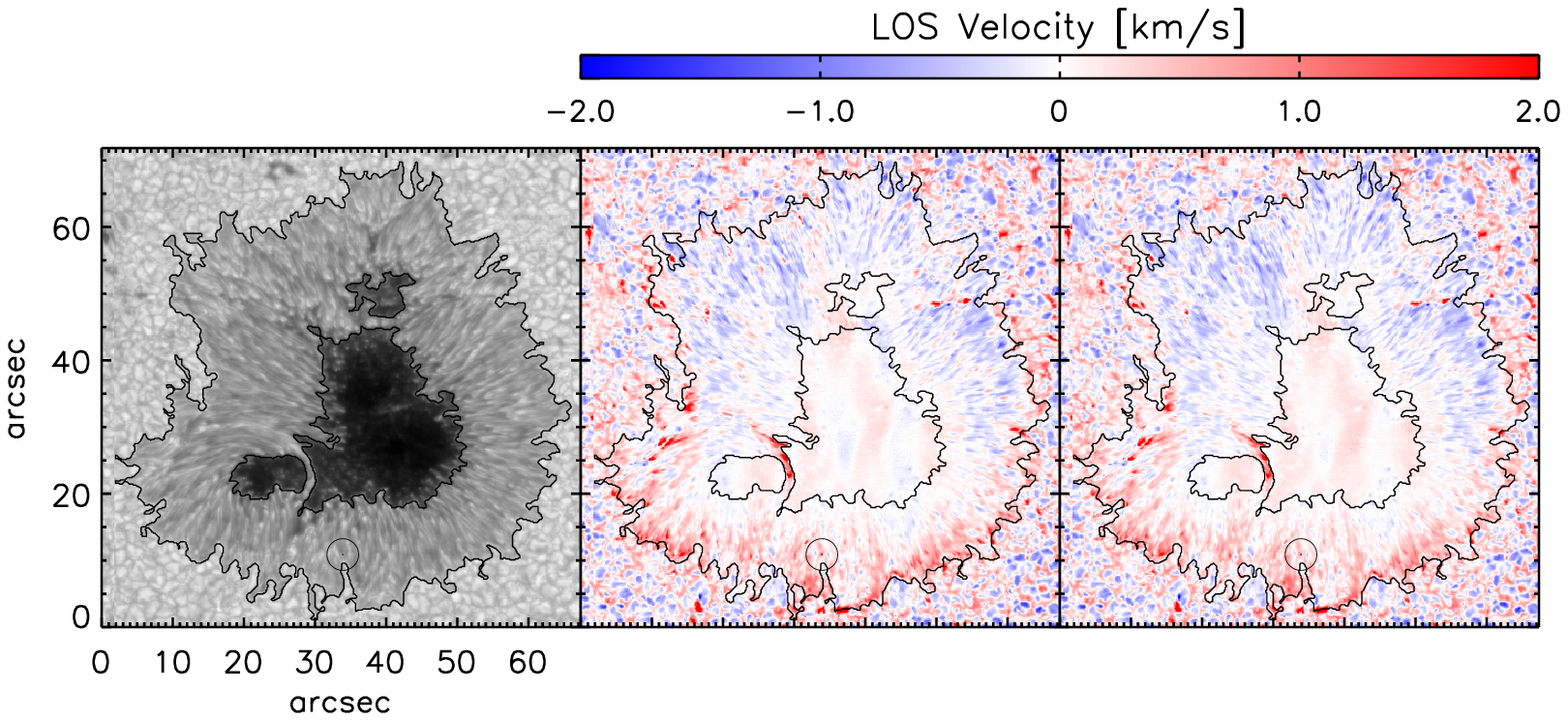}
}
\vspace{-140pt}
\caption{Leading sunspot in NOAA AR 10953 on 2007 May 01 as seen in continuum intensity (left), LOS velocity
derived from the \ionn{Fe}{1}\,6301.5\,\AA\, line (middle) and \ionn{Fe}{1}\,6302.5\,\AA\, line (right). The black circle
at ($x$, $y$) = (34\arcsec, 11\arcsec) encloses the pixel (black dot) whose profile is shown in red in Fig.~\ref{fig05} and represents
the EF returning to the photosphere.}
\label{fig01}
\end{figure*}

\section{Data Analysis}
\label{analysis}
\subsection{Pre-processing}
\label{invert}
The Level-1 Stokes profiles were normalized by the QS continuum intensity as an average
of all pixels where the polarization signal was less than three times the noise level.
This also yielded the average QS profile which was used to calculate the reference wavelengths
for the two \ionn{Fe}{1} lines separately. The 112-pixel wavelength window was then split into
two, of 56 wavelength points each, and the corresponding profiles of the two lines were 
then inverted independently using the 
parallel version of the SIR code \citep{2015IAUS..305..251T}. This was done so that 
the clustering was performed on the basis of the shape of both \ionn{Fe}{1} lines.
We ran a single cycle of SIR 
with height-independent LOS velocity and the vector magnetic field, along with two nodes for 
temperature. Figure~\ref{fig01} shows the sunspot in continuum intensity as well as the LOS velocity
derived from the inversions for the two \ionn{Fe}{1} lines.
The resultant LOS velocity was then used to shift the Stokes \textit{V} profile 
in wavelength from which the central 44 wavelength points were used. Finally, the shifted spectra 
of both \ionn{Fe}{1} lines were combined to yield 88 points in wavelength that were then normalized 
to the maximum unsigned value. The last two steps are similar to the method described in 
\citet{2011A&A...530A..14V}.
The clustering is done for all pixels in the sunspot excluding
only those in the three umbral cores, which resulted in the classification of 93458 pixels.
The penumbral region includes a granular and filamentary LB located in the northern and southern part
of the sunspot, respectively. We note that as the clustering scheme works exclusively on 
the shape of the Stokes $V$ profile, the effect of normalization by the individual maxima  
masks the actual amplitude of the spectra. The verification of the final output was done by 
inspecting the minor clusters (the ones with a small number of samples) and comparing their true 
amplitude to the common and dominant profiles present in the penumbra (Sect.~\ref{spectra}). 

\subsection{k-means Clustering}
\label{kmeans}
In order to group the Stokes \textit{V} profile based on their shapes, we use the {\tt{k-means}} 
clustering technique \citep{Loyd1982,Forgy1965}\footnote{Technique first proposed by Lloyd in 1957}.
This method is widely used in machine learning, data mining etc. where $N$ data points are grouped
into $k$ clusters. This is done by using a set of $k$ cluster centers at random from the $N$ data points
and measuring the Euclidean distance of each data point from the cluster centers. The data point 
is assigned the index of that cluster center to which the distance is a minimum. 
When the indexing of all data points is completed, the new cluster centers are computed 
for each resulting cluster as the average of the samples (elements) in each cluster. 
The process of assigning a data point 
to a cluster center is complete when the centroids of each cluster along with the variance remain 
unchanged. 

While {\tt{k-means}} is quite simple to implement, the clustering result depends on 
the choice of the initial cluster centers as well as the number of clusters. In the classical approach, 
the clustering scheme is run several times with a different random initialization and
the resulting variance and cluster center from each run are then analyzed. 
Since each sample or spectra is assigned an index based on its least distance from all the 
cluster centers, we computed the initial cluster centers in the following manner. 
The first cluster center corresponds to the profiles that
is definitely a cluster class, namely nearly anti-symmetric Stokes \textit{V} profiles. The choice 
of the above was done from the pixel where the corresponding Stokes \textit{V} profile had the least
area asymmetry. The second cluster center was determined as that profile which had a maximum 
distance to the first cluster center. Once the first and second initial cluster centers were 
determined, the same process was repeated for the remaining cluster centers, such that 
the distance of the next cluster center was the maximum of the total distance with the previous 
cluster centers while also ensuring that the data points corresponding to the previous cluster centers 
were excluded from the calculation. The rationale behind this approach is to start off with cluster centers
that are spaced out from each other as far as possible.

\begin{figure}[!h]
\centerline{
\includegraphics[angle=90,width = \columnwidth]{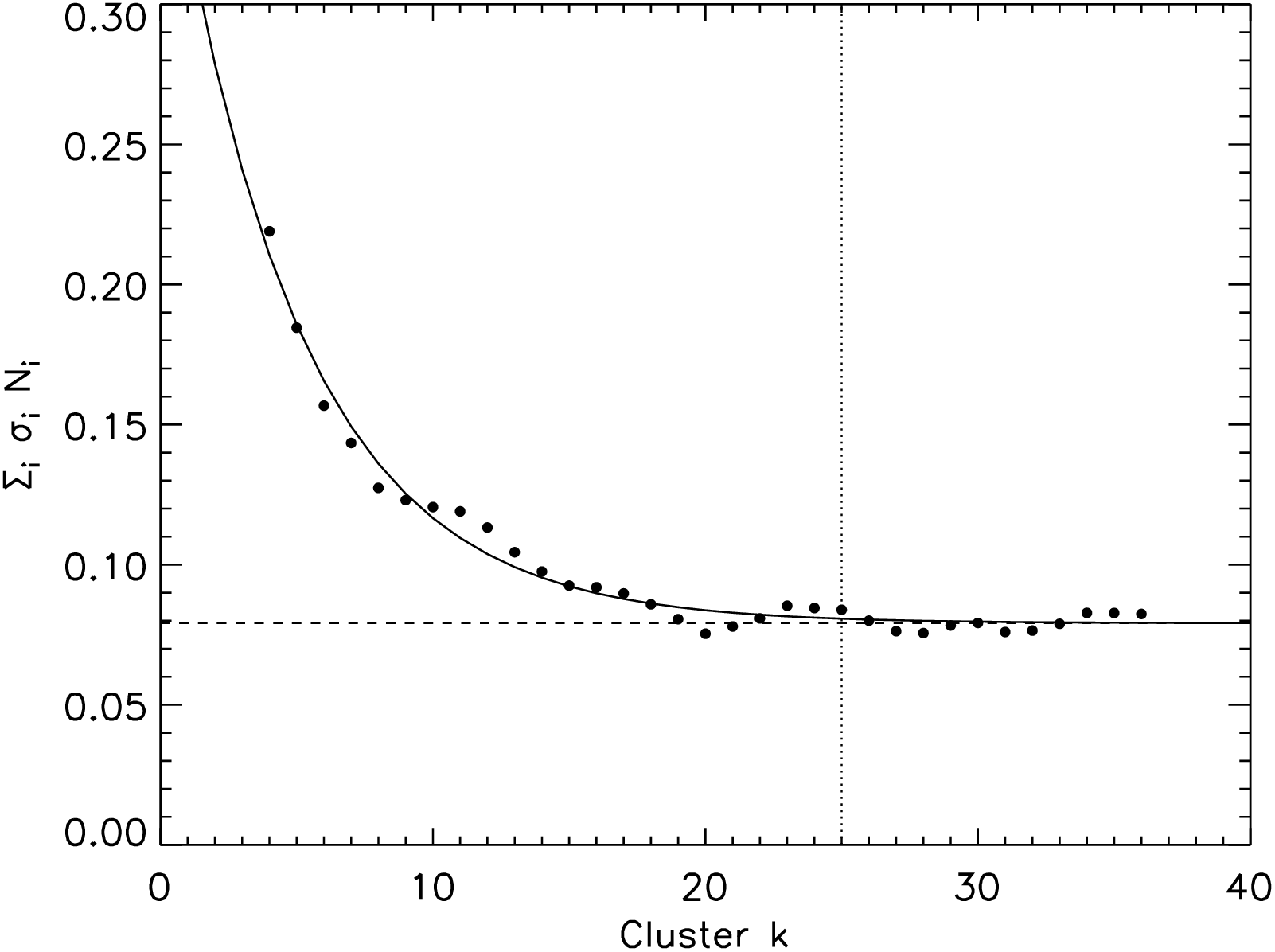}
}
\vspace{0pt}
\caption{Variation of the standard deviation weighted by the number of profiles in a given 
cluster for different clusters $k$.
The summation in the $y-$axis is performed over the index $i$ for those clusters where the mean Stokes $V$ profile
is nearly anti-symmetric. The solid line is an exponential decaying function fitted to the data points (black
filled circles). The dashed line indicates where the quantity on the $y-$axis has settled despite an 
increase in $k$. This yields the optimum value of {\tt{k}} $=25$ as shown with the vertical dotted line.}
\label{fig02}
\end{figure}

\begin{figure*}[!ht]
\centerline{
\hspace{25pt}
\includegraphics[angle=90,width = 0.43\textwidth]{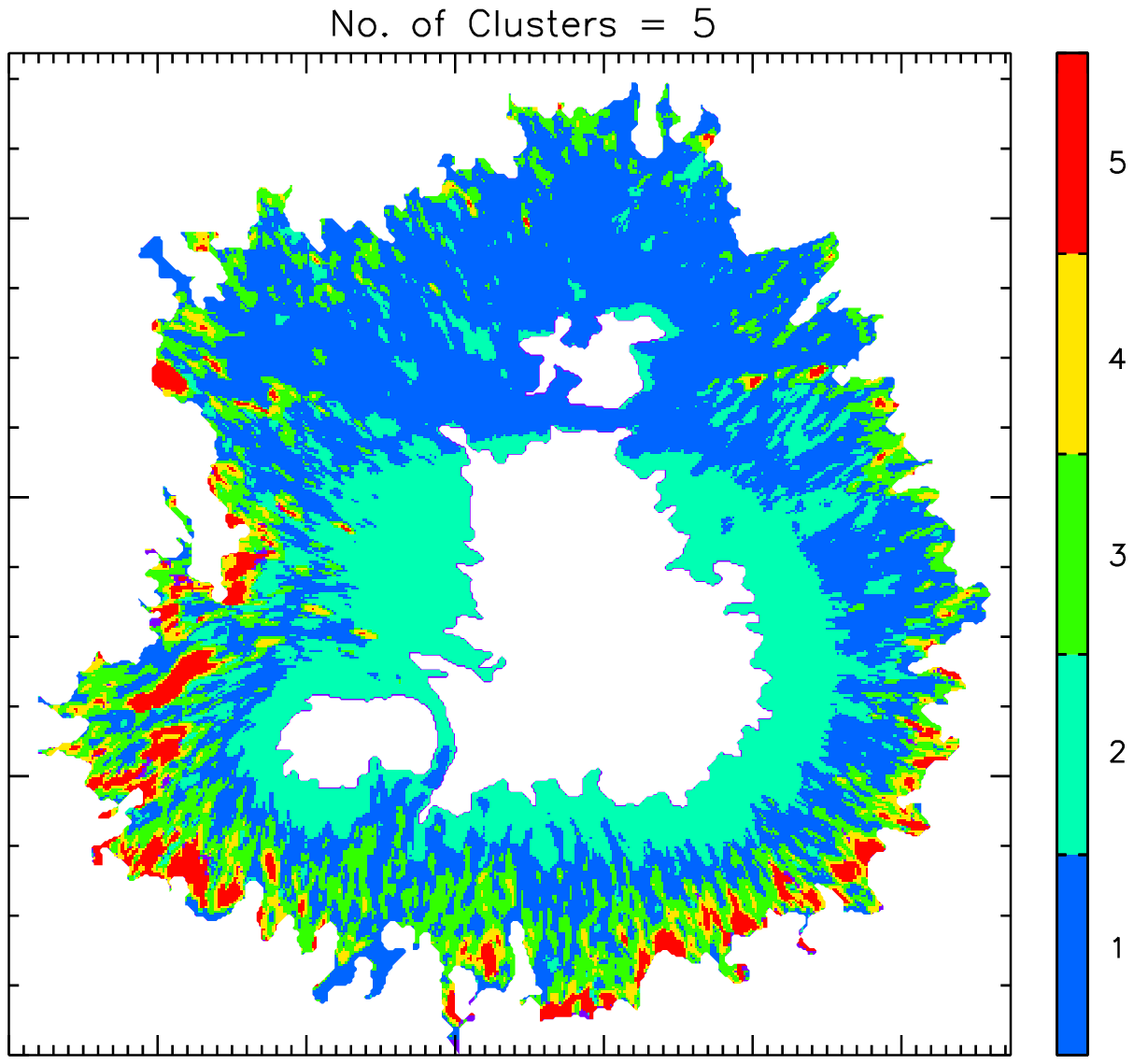}
\hspace{-60pt}
\includegraphics[angle=90,width = 0.43\textwidth]{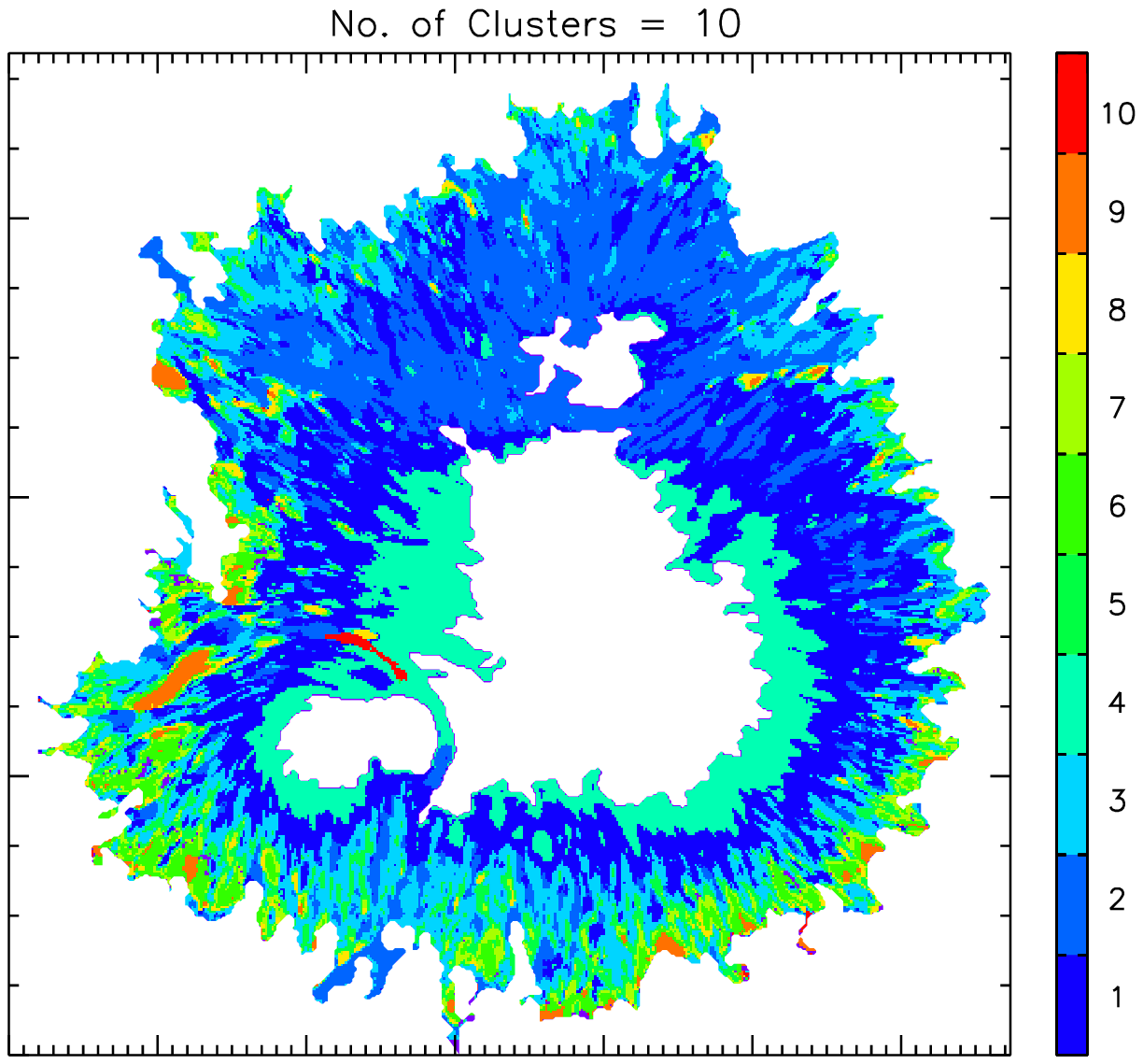}
\hspace{-60pt}
\includegraphics[angle=90,width = 0.43\textwidth]{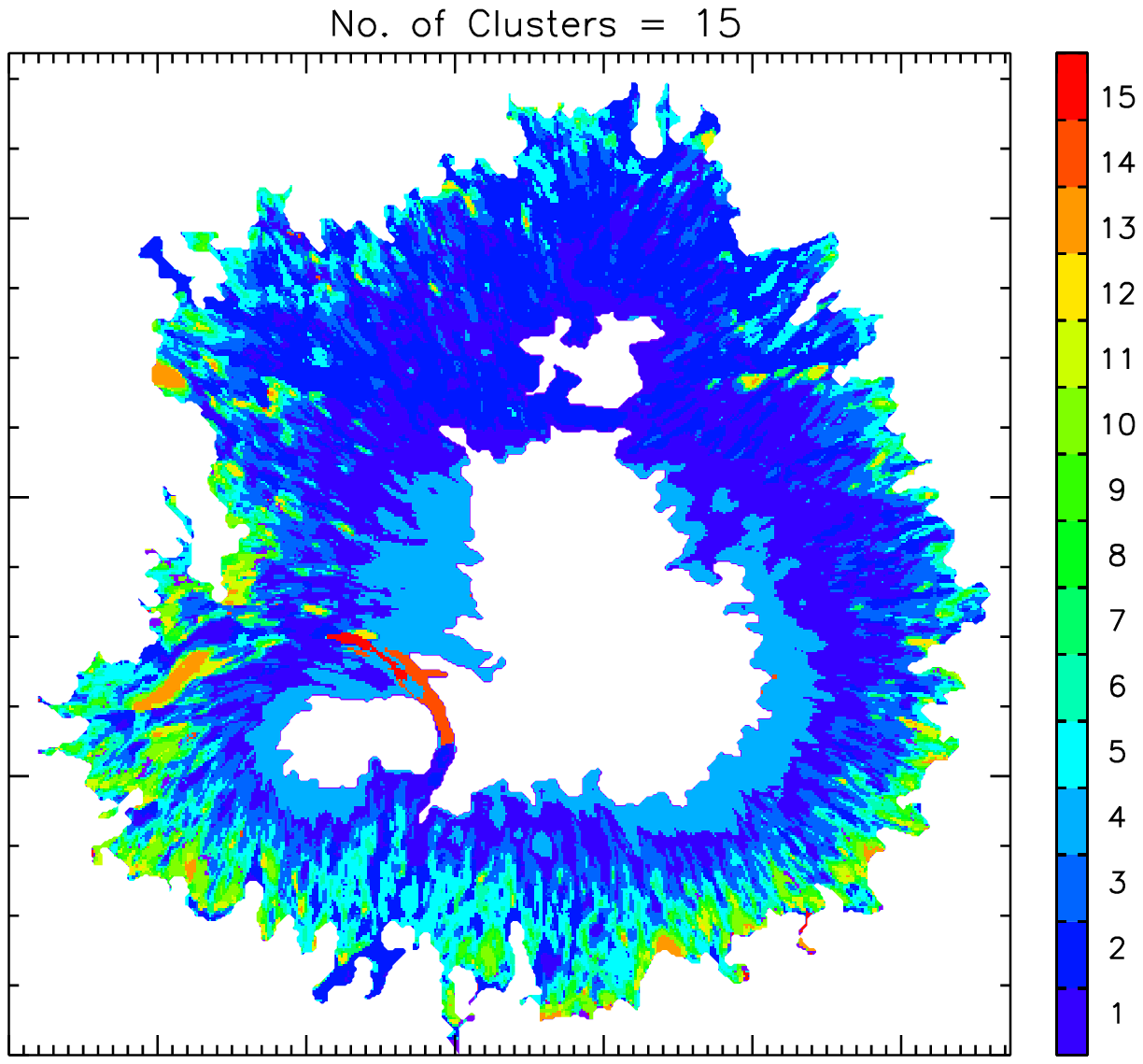}
}
\centerline{
\hspace{25pt}
\includegraphics[angle=90,width = 0.43\textwidth]{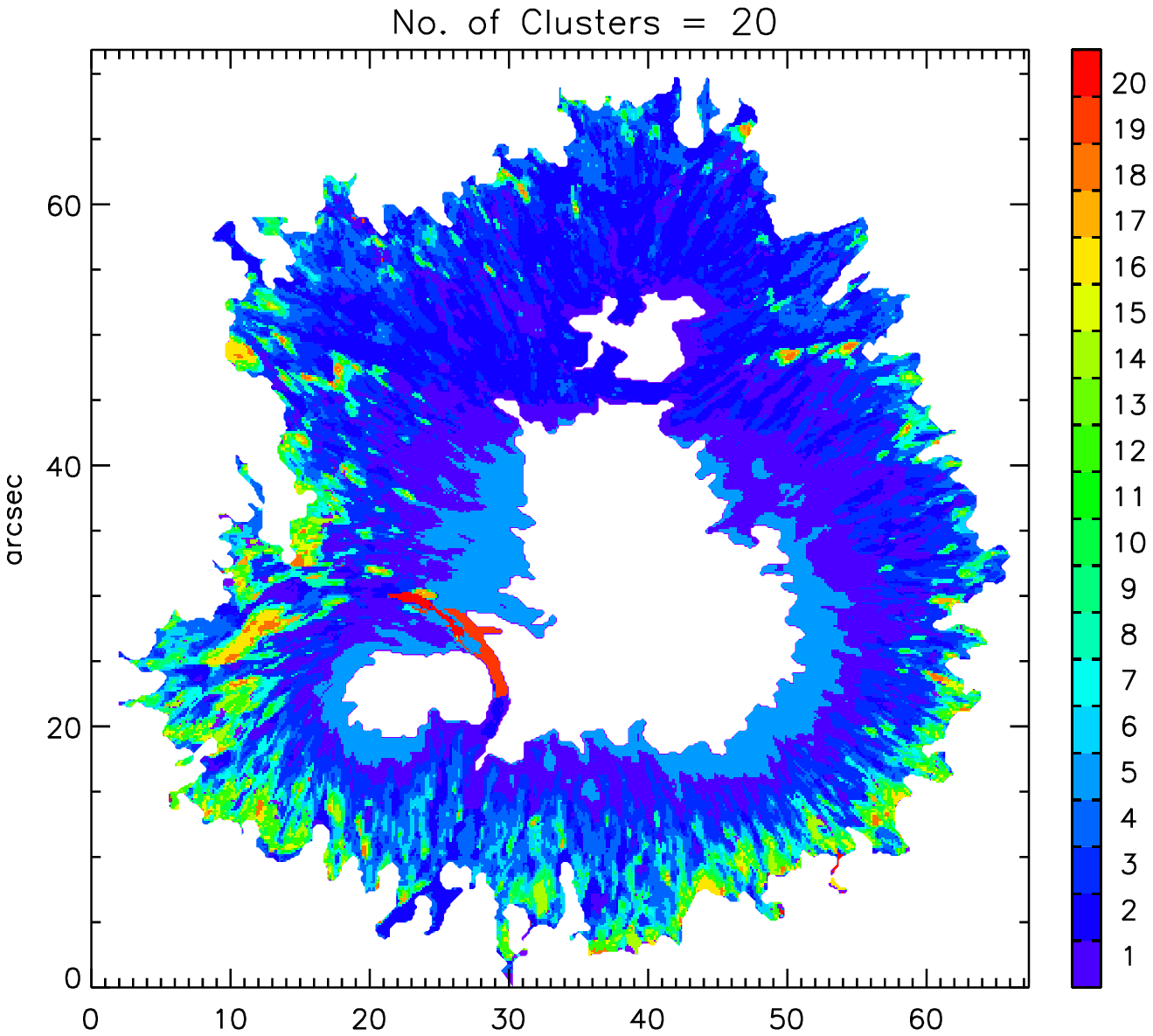}
\hspace{-60pt}
\includegraphics[angle=90,width = 0.43\textwidth]{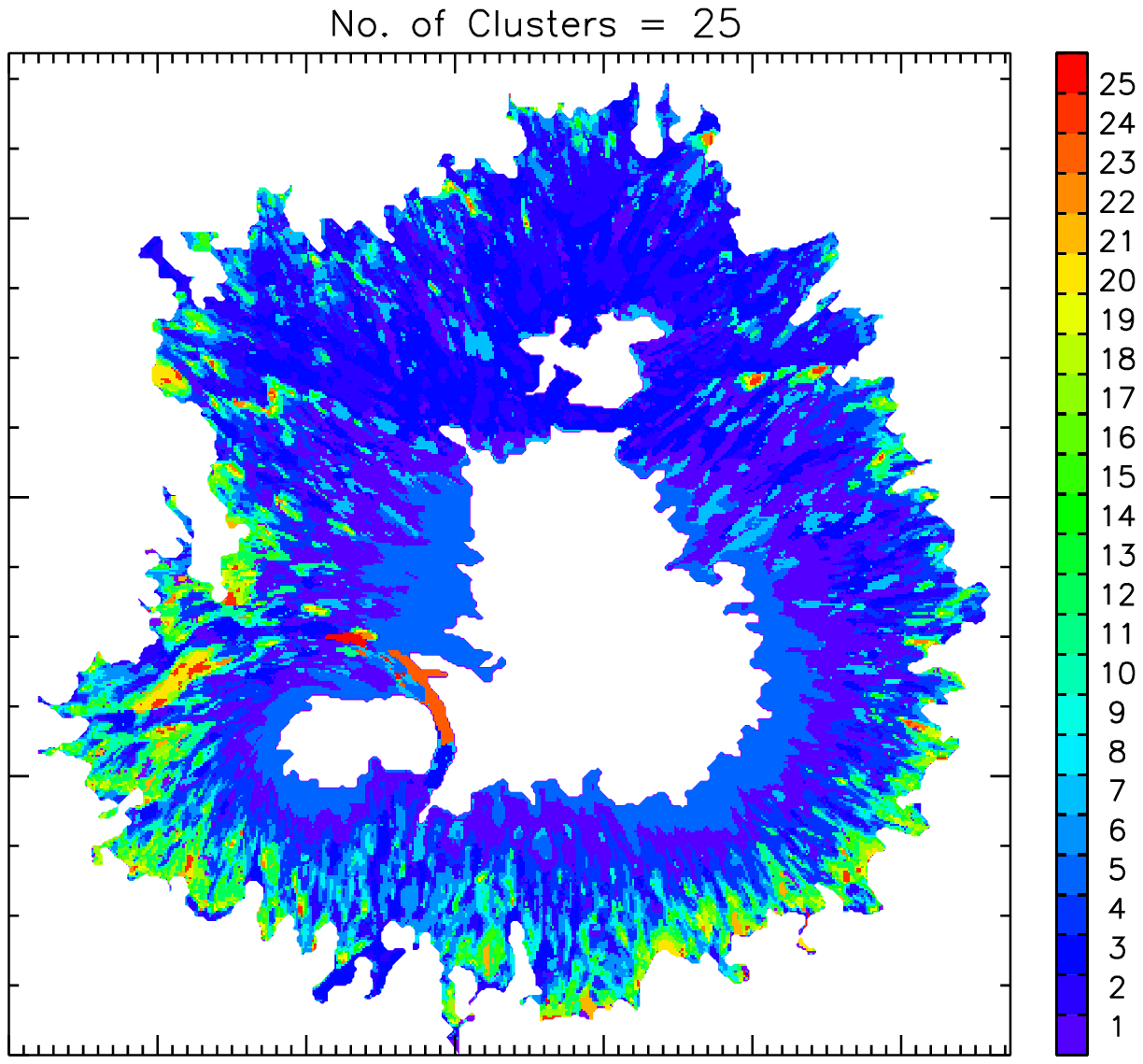}
\hspace{-60pt}
\includegraphics[angle=90,width = 0.43\textwidth]{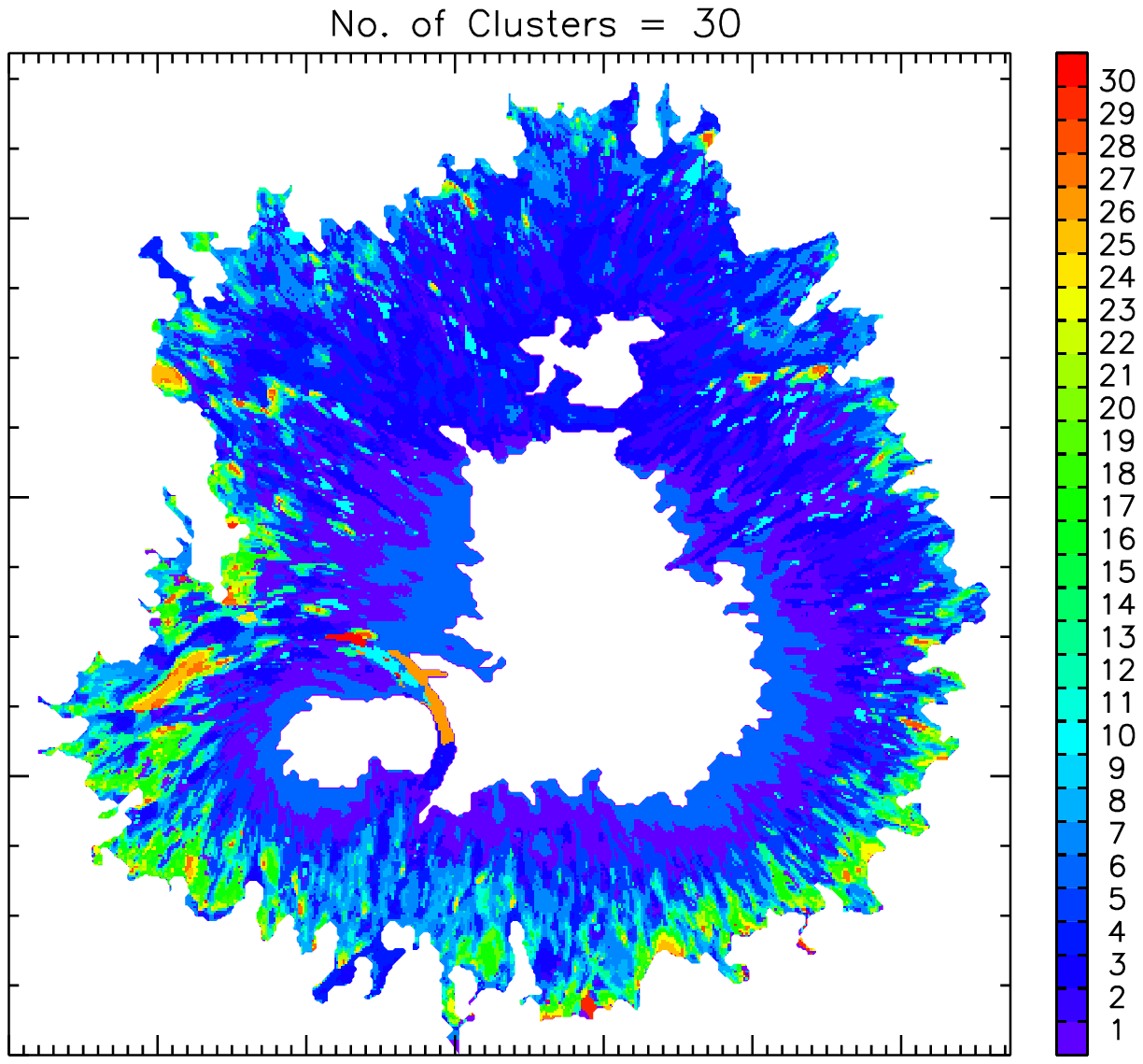}
}
\vspace{-5pt}
\caption{Inter-cluster variation for select number of clusters. The numbers in the 
legend are in ascending order from the most populous (violet) cluster to the most 
scarce (red) cluster.}
\label{fig03}
\end{figure*}

\begin{figure*}[!ht]
\centerline{
\hspace{50pt}
\includegraphics[angle=90,width = 0.45\textwidth]{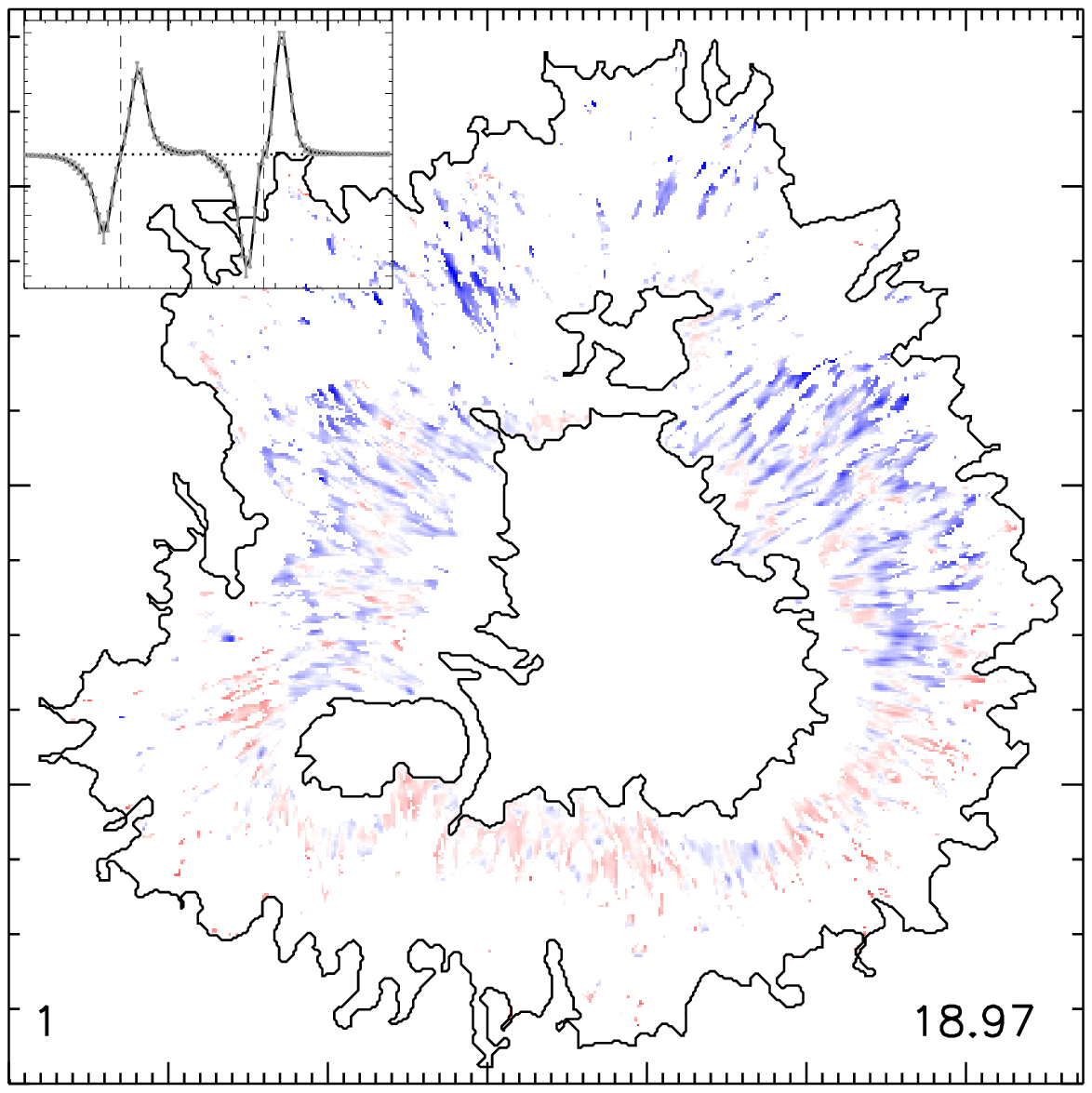}
\hspace{-70pt}
\includegraphics[angle=90,width = 0.45\textwidth]{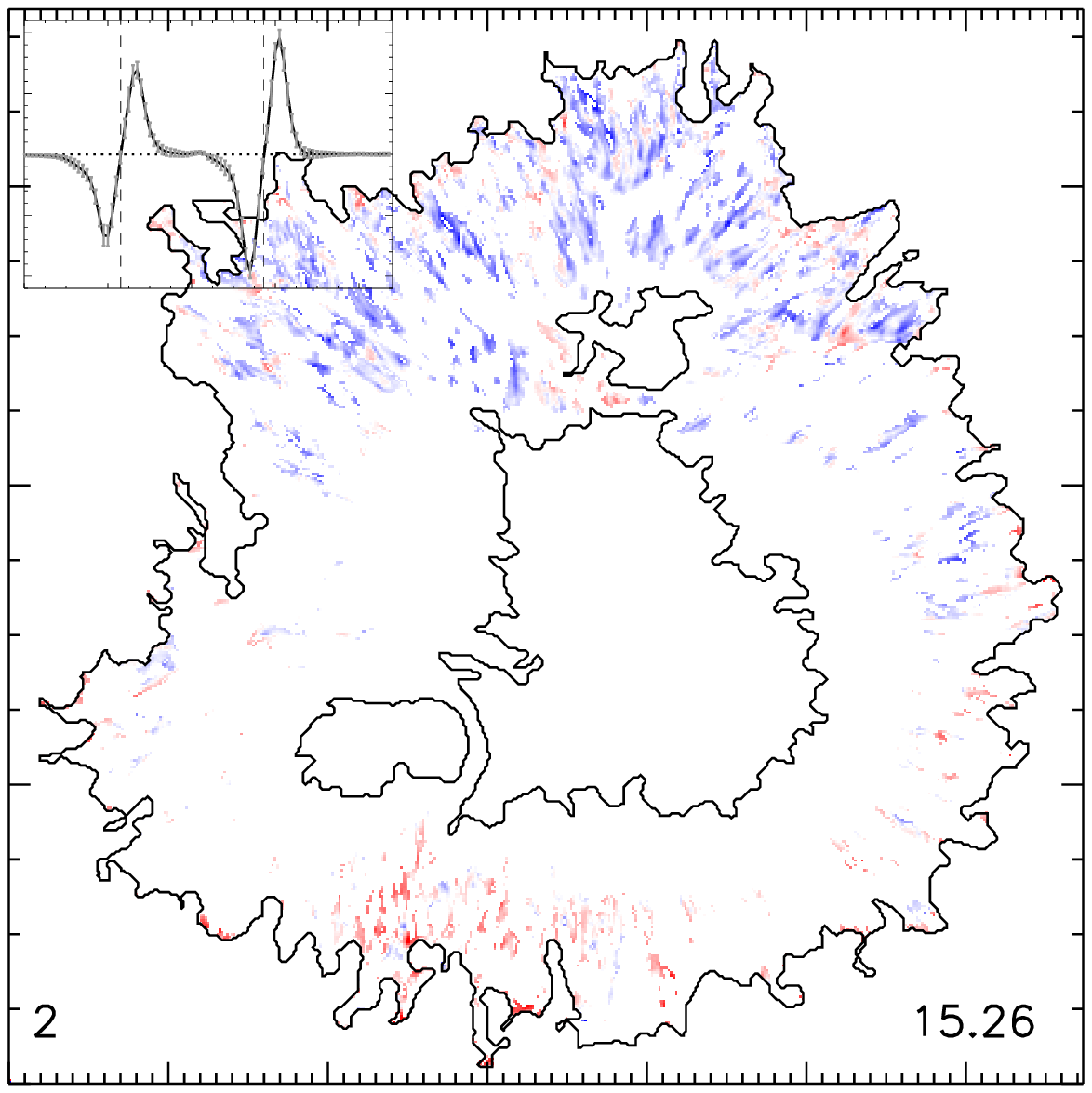}
\hspace{-70pt}
\includegraphics[angle=90,width = 0.45\textwidth]{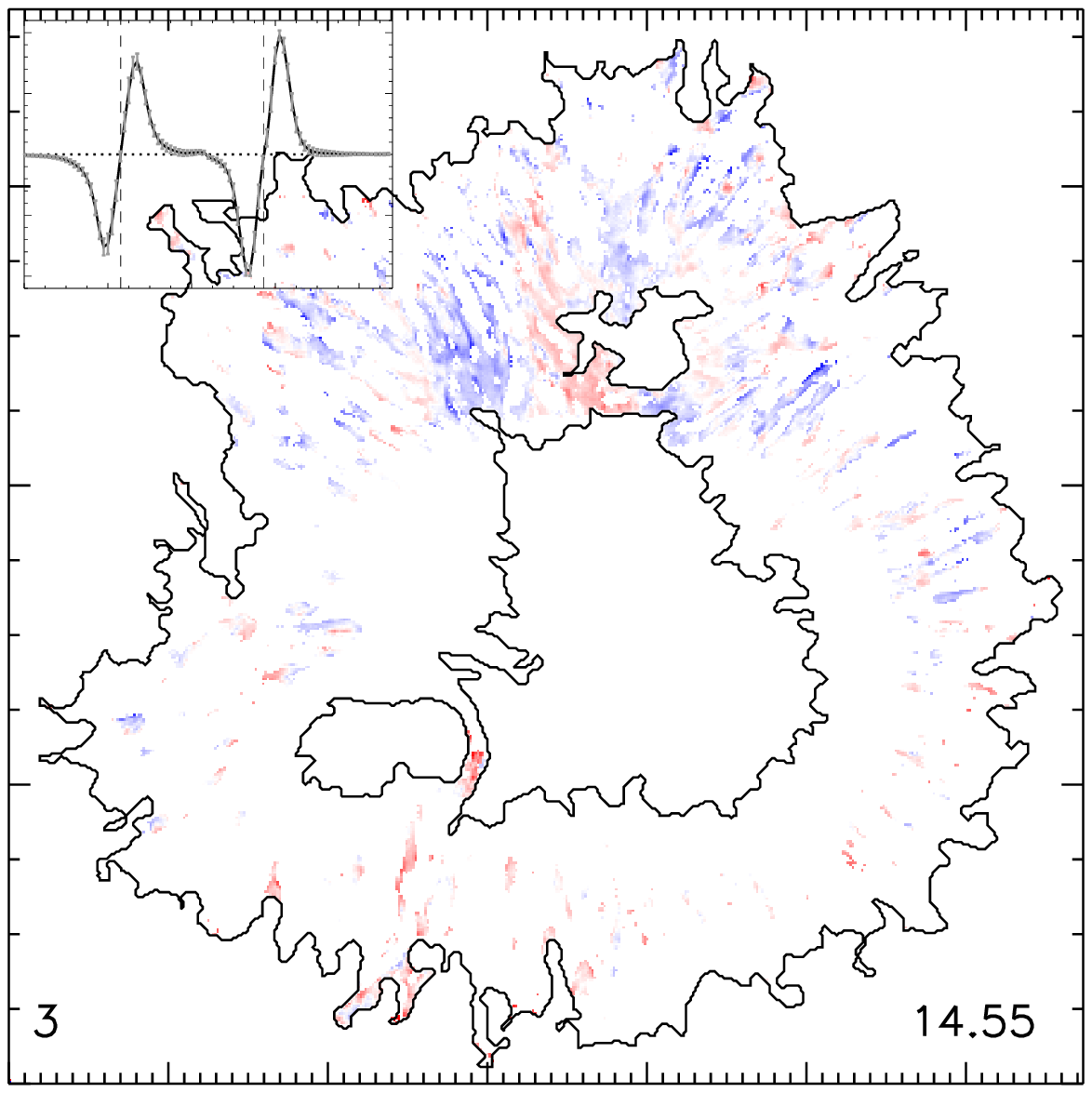}
}
\centerline{
\hspace{50pt}
\includegraphics[angle=90,width = 0.45\textwidth]{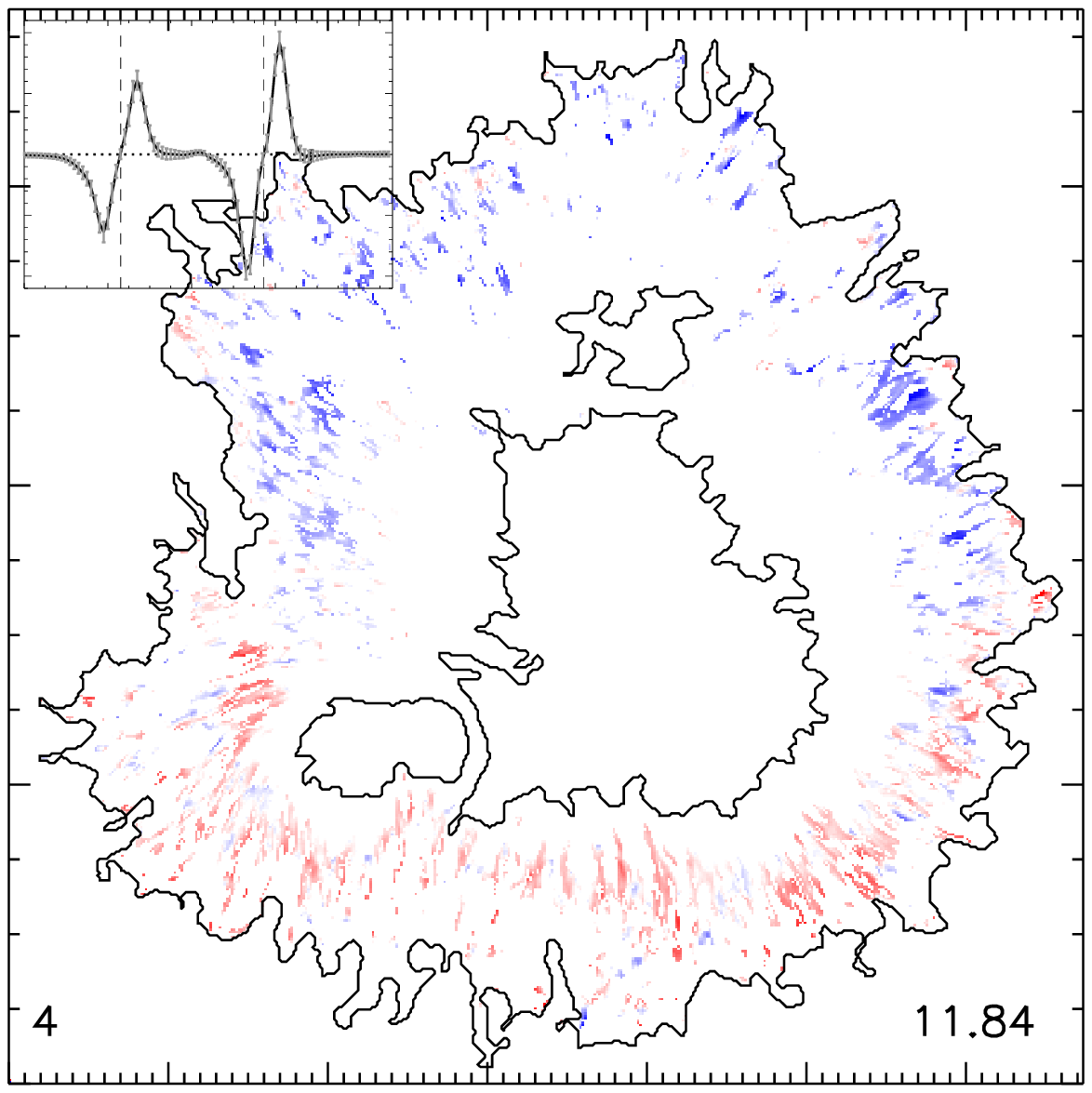}
\hspace{-70pt}
\includegraphics[angle=90,width = 0.45\textwidth]{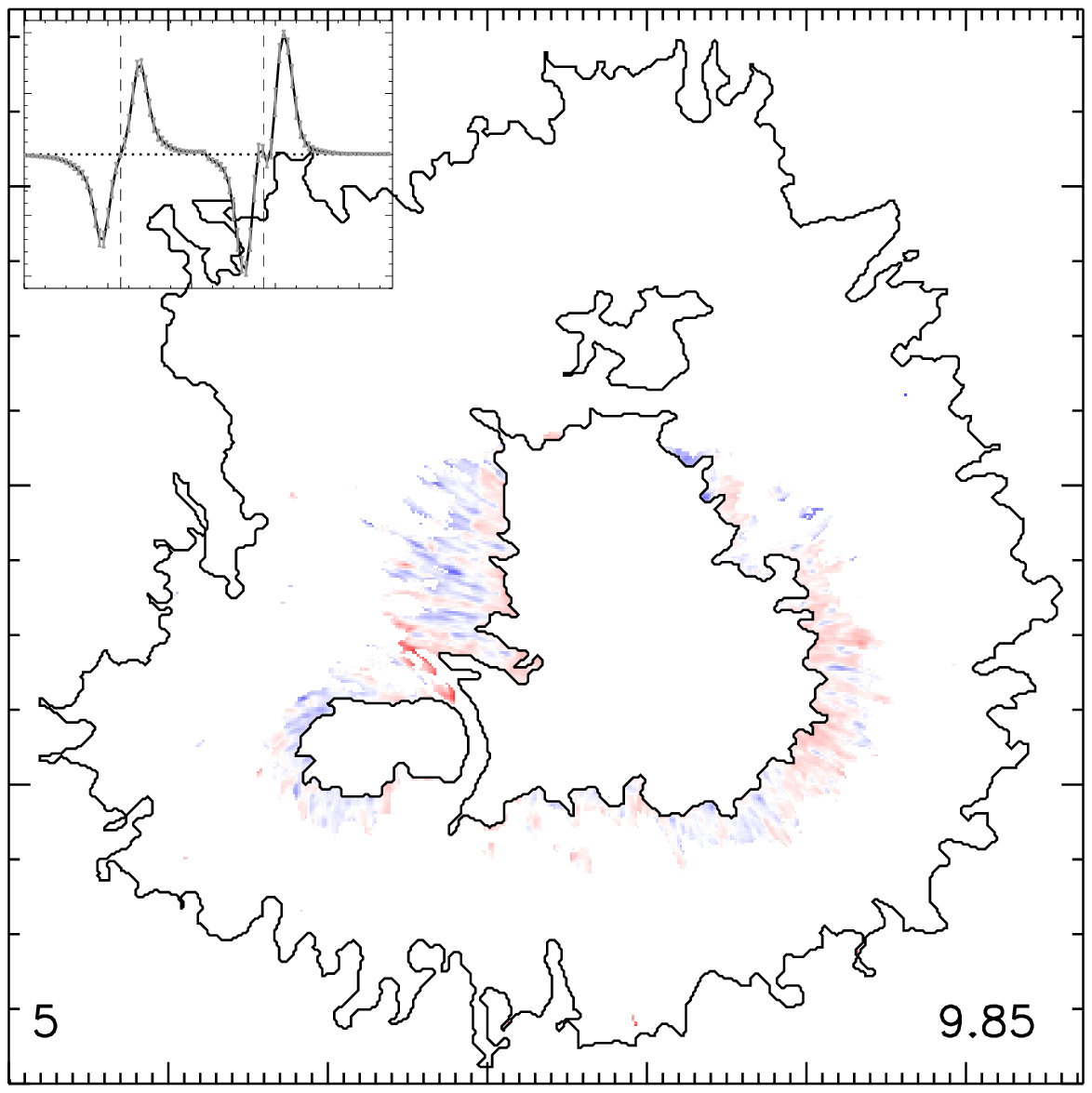}
\hspace{-70pt}
\includegraphics[angle=90,width = 0.45\textwidth]{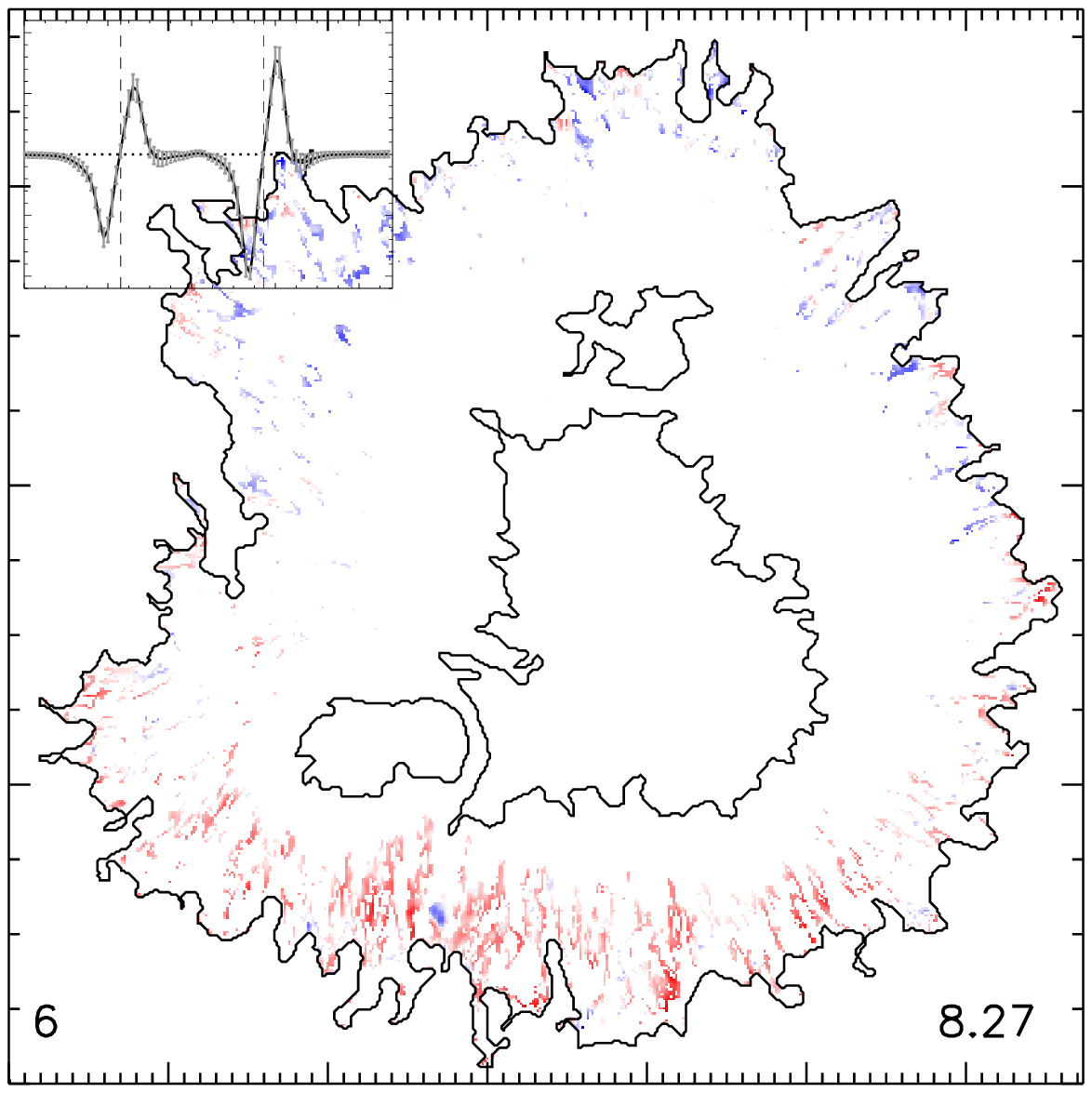}
}
\centerline{
\hspace{50pt}
\includegraphics[angle=90,width = 0.45\textwidth]{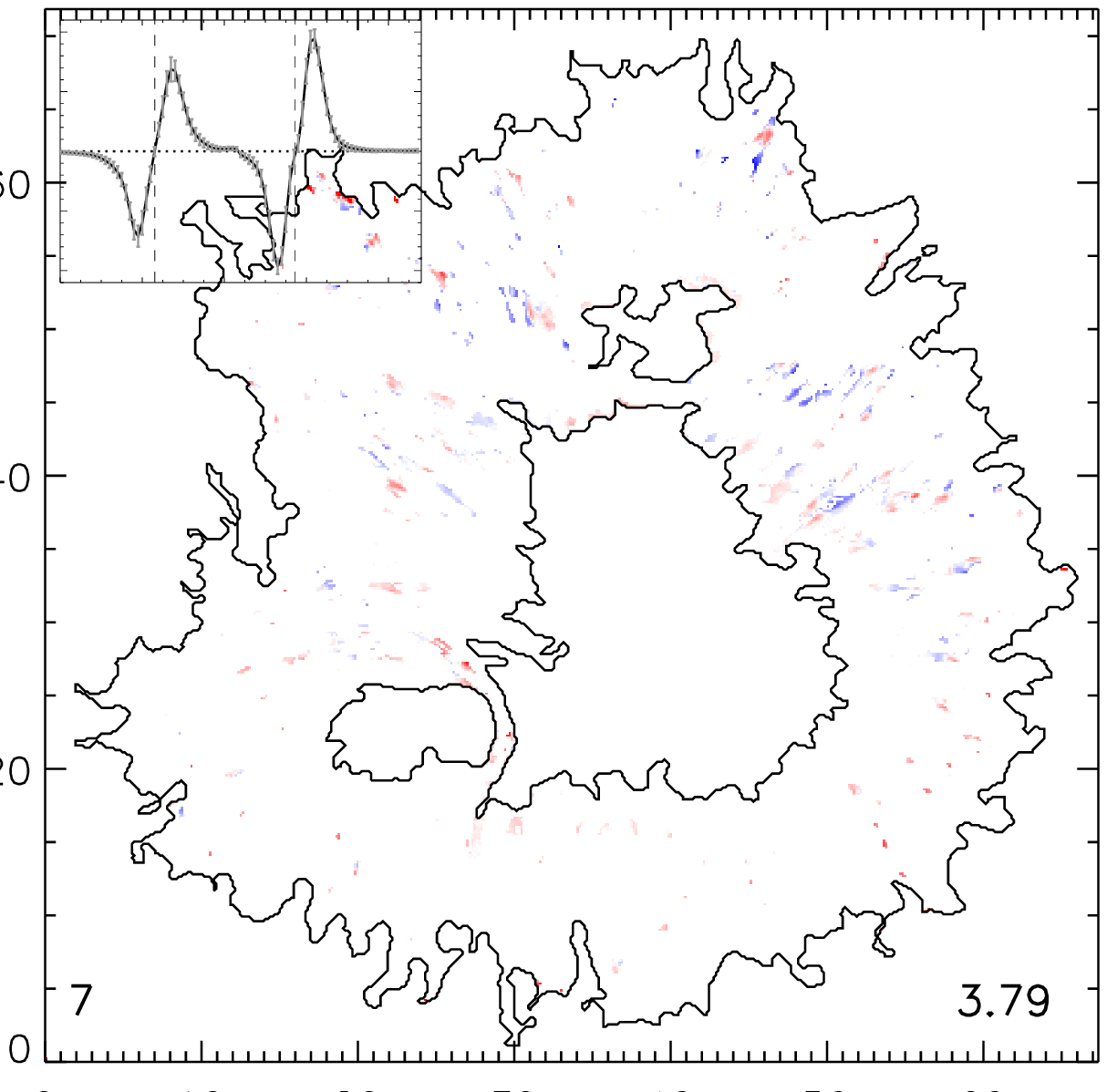}
\hspace{-70pt}
\includegraphics[angle=90,width = 0.45\textwidth]{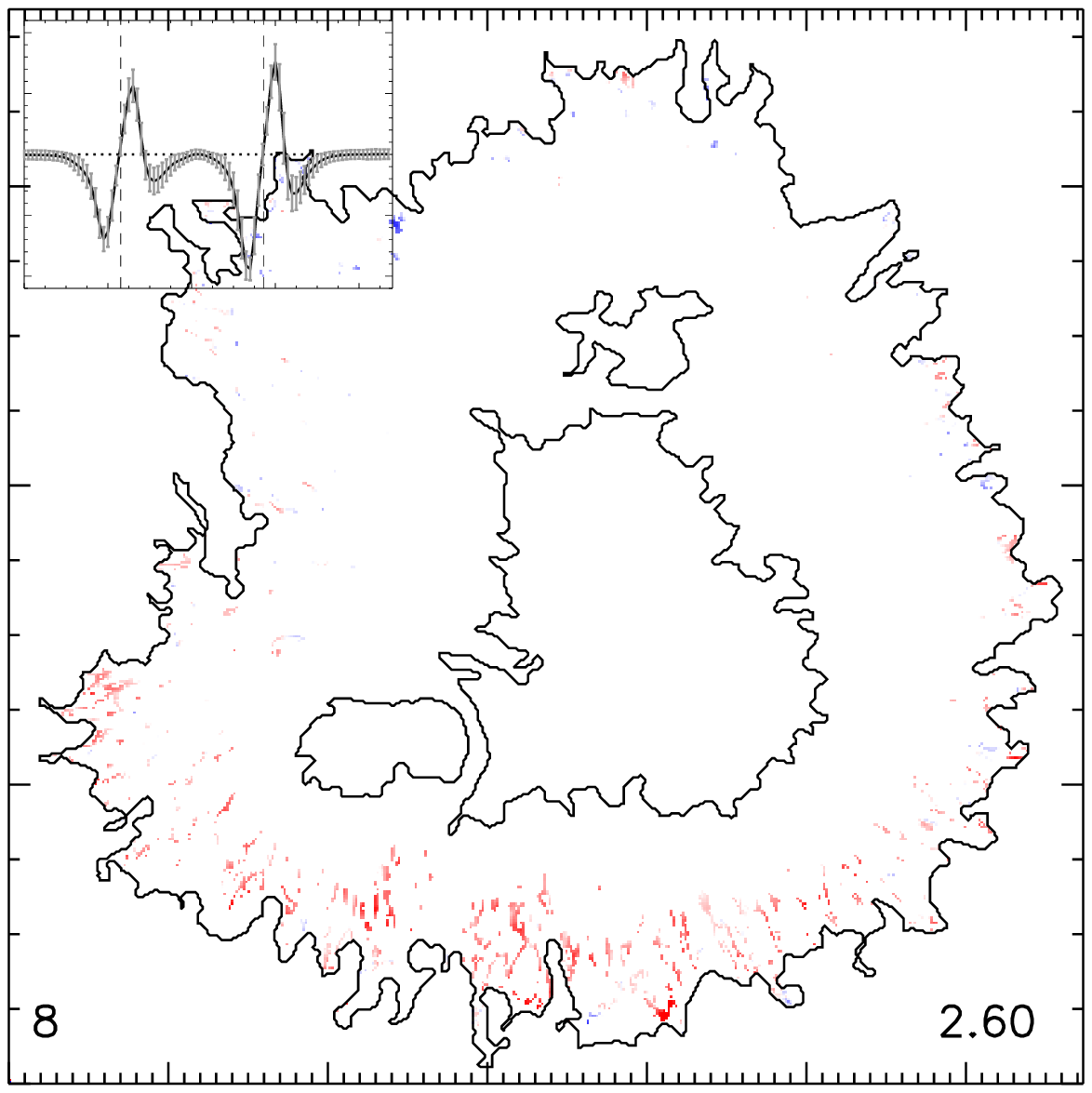}
\hspace{-70pt}
\includegraphics[angle=90,width = 0.45\textwidth]{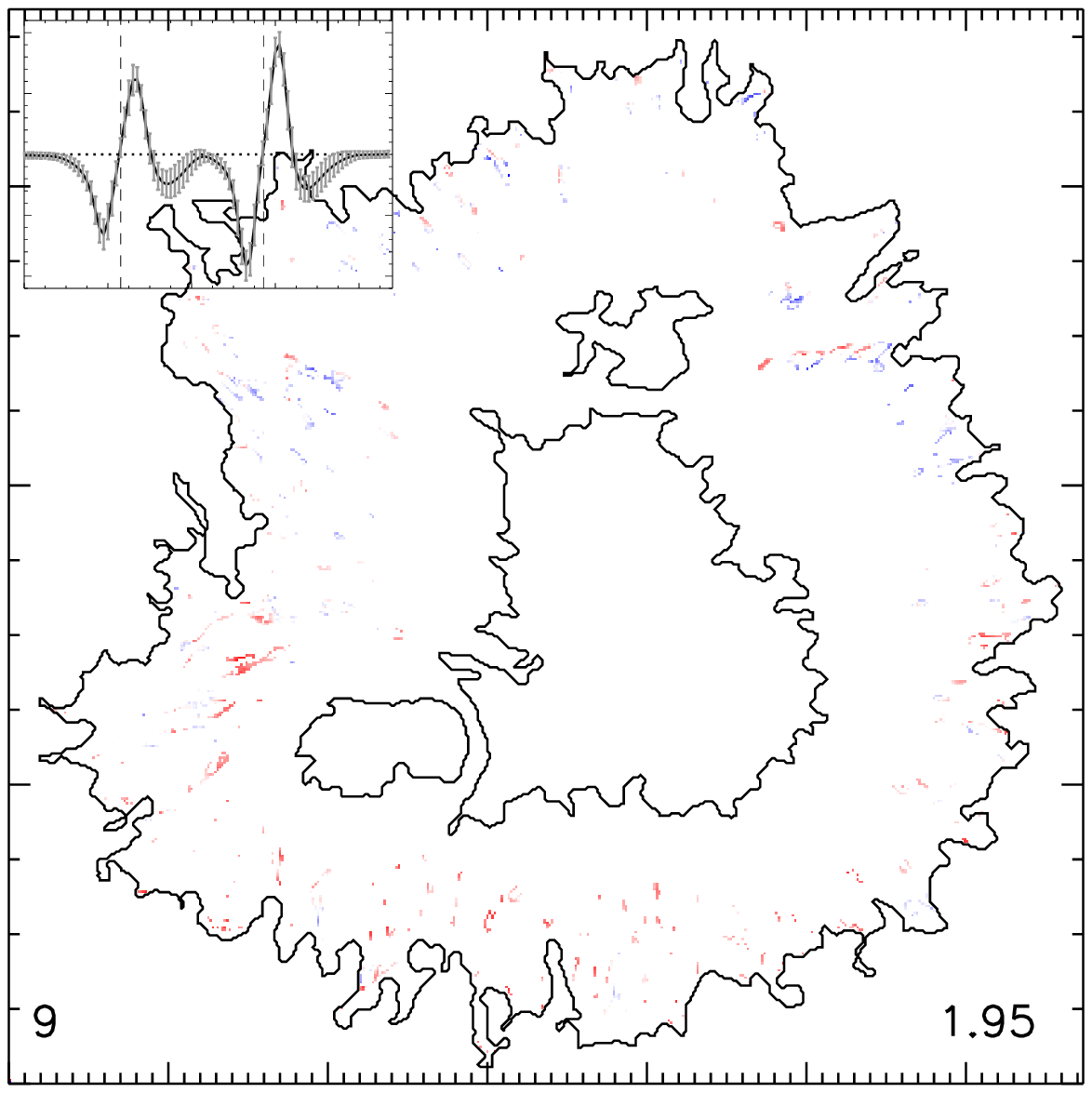}
}
\vspace{5pt}
\caption{Spatial distribution of individual clusters in the penumbra for the first nine clusters when using 
a total of 25 clusters. The corresponding LOS velocity map in the background at the location of each cluster
has been scaled between $\pm 1$\,km\,s$^{-1}$. The mean Stokes $V$ profile for each cluster is shown in the top 
left corner with the grey error bars signifying a 1\,$\sigma$ variation over the mean. The two vertical 
dashed lines represent the Stokes $V$ zero crossing position of the two \ionn{Fe}{1} lines. The horizontal, 
dotted line corresponds to an amplitude of zero. The numbers in the bottom corners represent the cluster number 
(left) and the percentage area occupied by that cluster in the penumbra (right).}
\label{fig04a}
\end{figure*}

\begin{figure*}[!ht]
\centerline{
\hspace{50pt}
\includegraphics[angle=90,width = 0.45\textwidth]{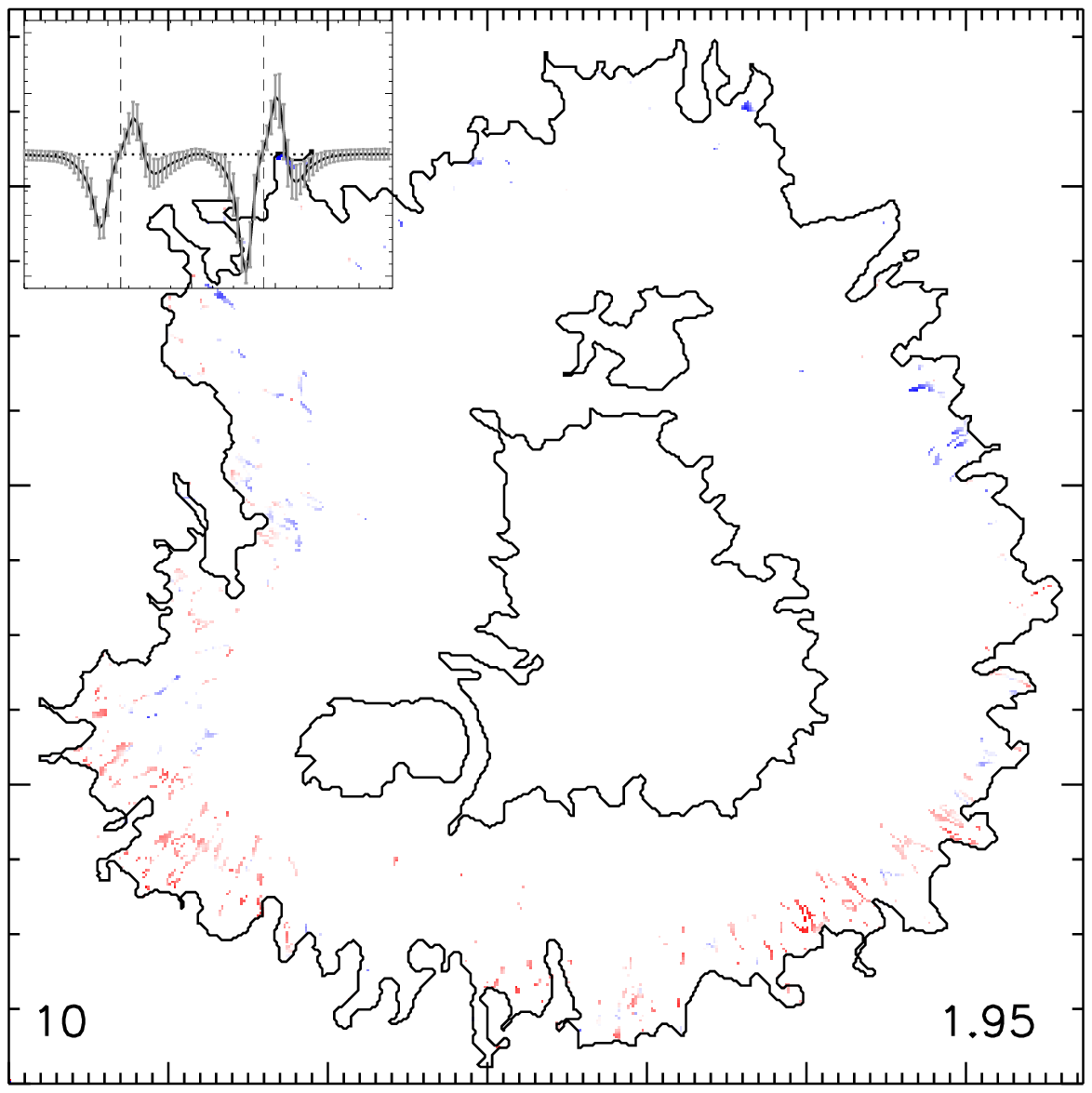}
\hspace{-70pt}
\includegraphics[angle=90,width = 0.45\textwidth]{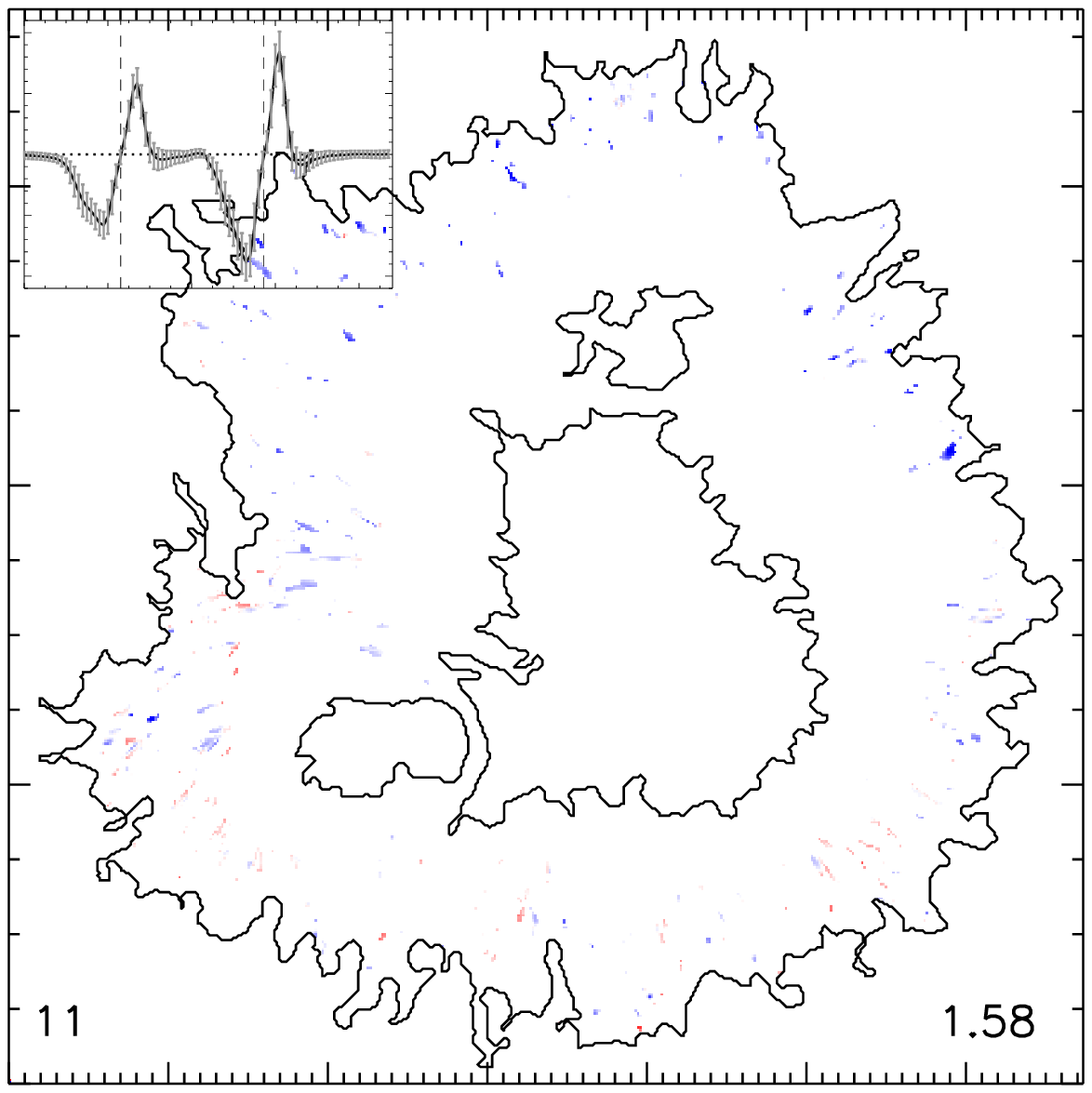}
\hspace{-70pt}
\includegraphics[angle=90,width = 0.45\textwidth]{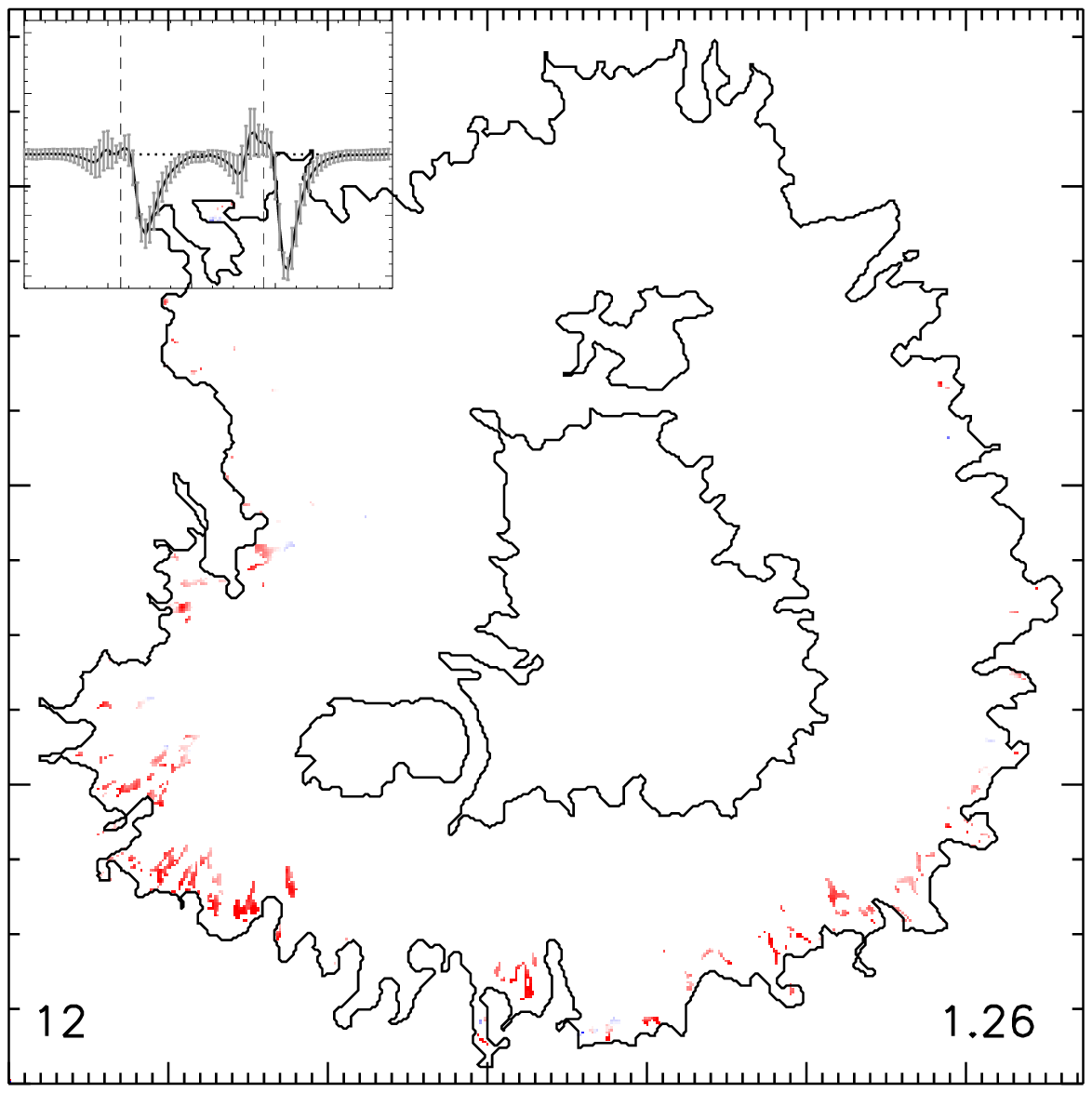}
}
\centerline{
\hspace{50pt}
\includegraphics[angle=90,width = 0.45\textwidth]{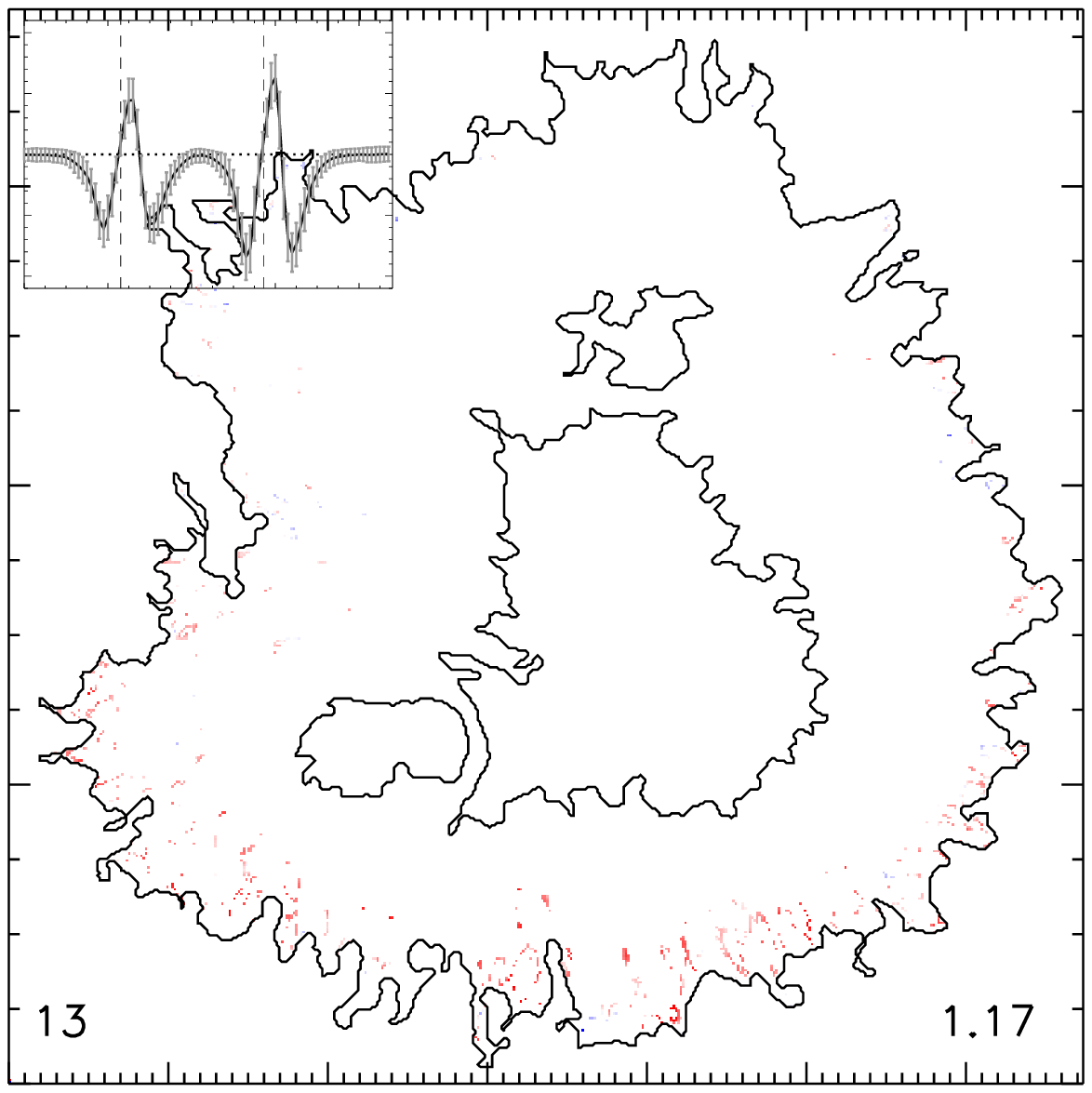}
\hspace{-70pt}
\includegraphics[angle=90,width = 0.45\textwidth]{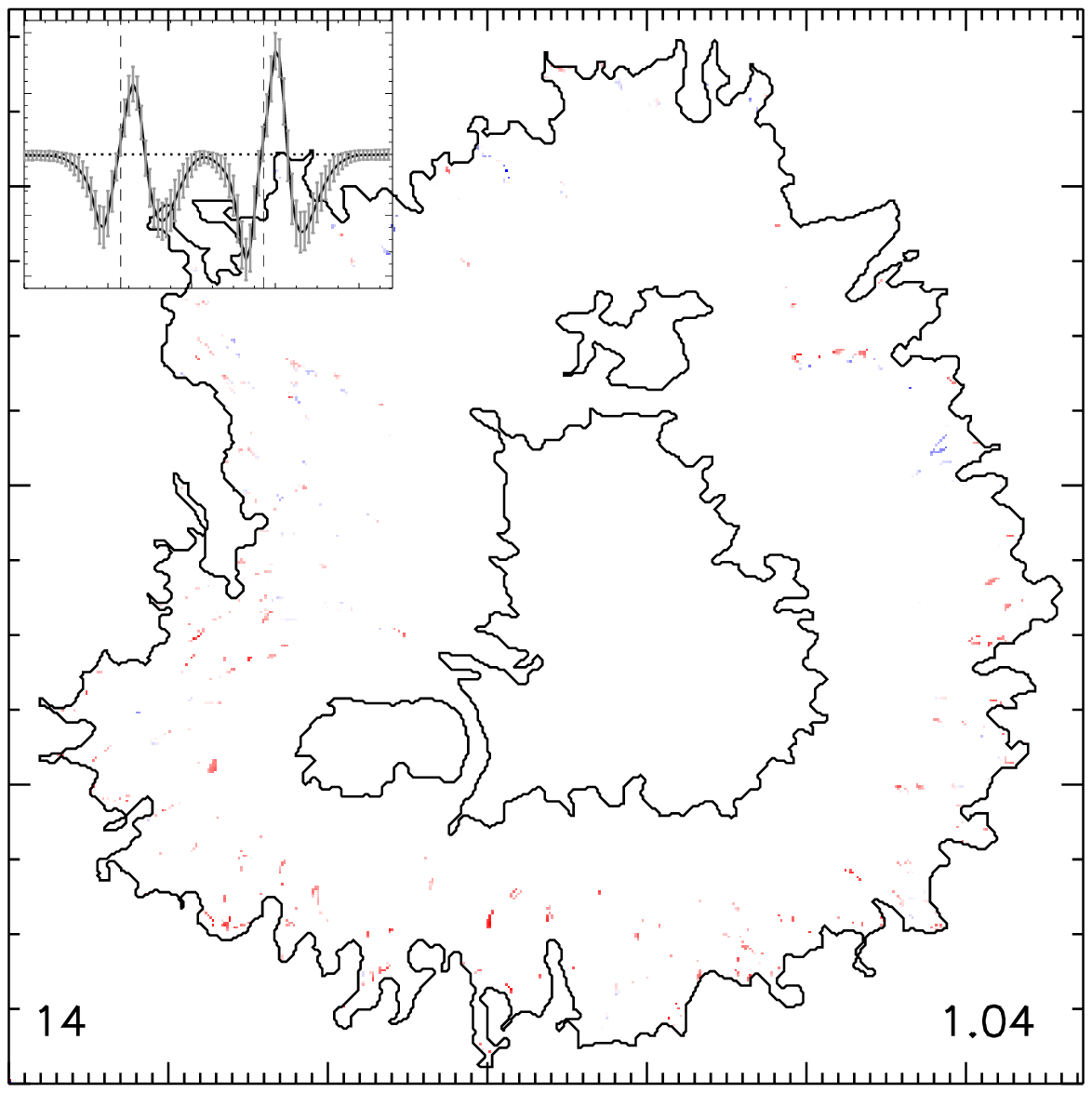}
\hspace{-70pt}
\includegraphics[angle=90,width = 0.45\textwidth]{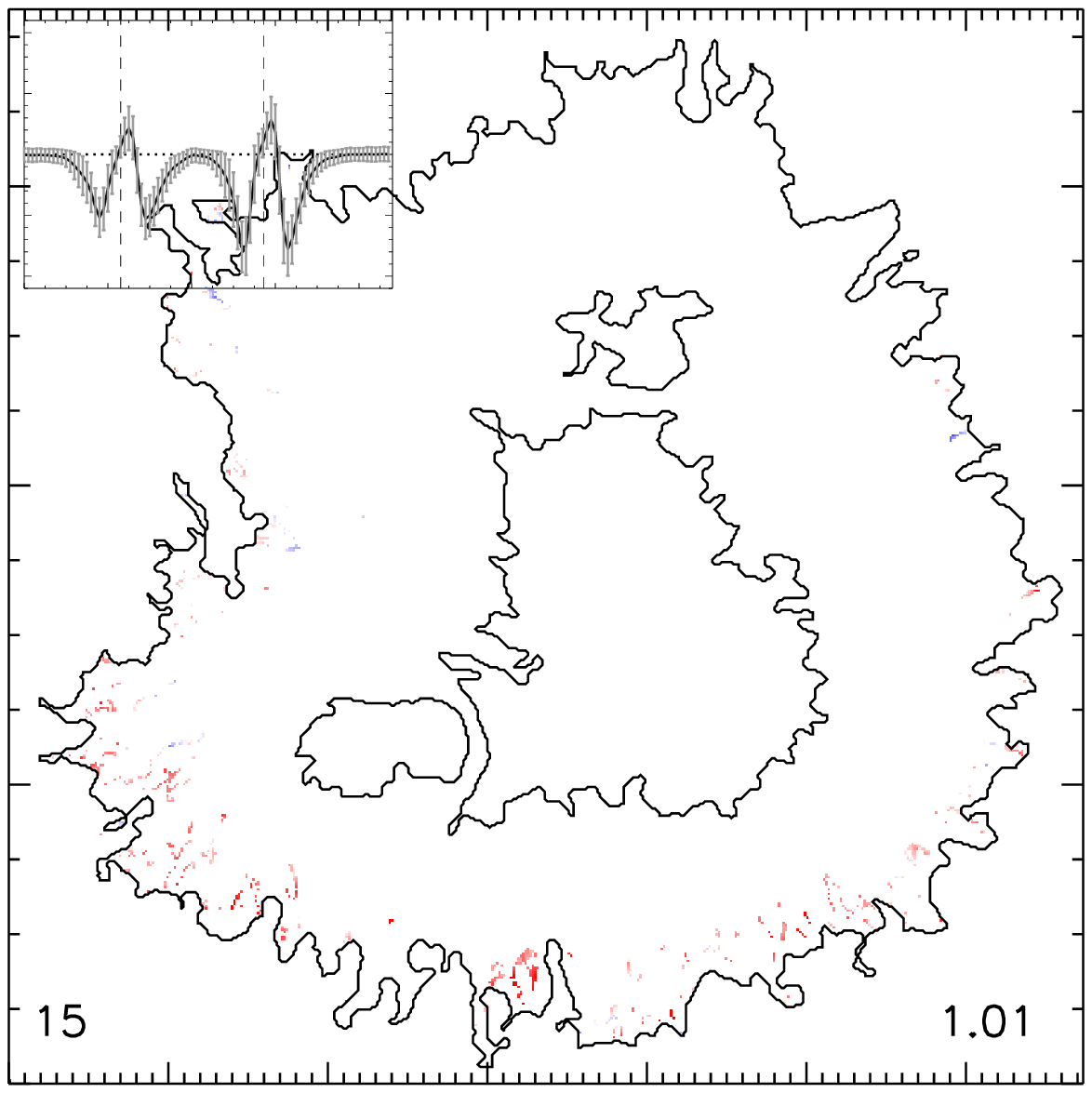}
}
\centerline{
\hspace{50pt}
\includegraphics[angle=90,width = 0.45\textwidth]{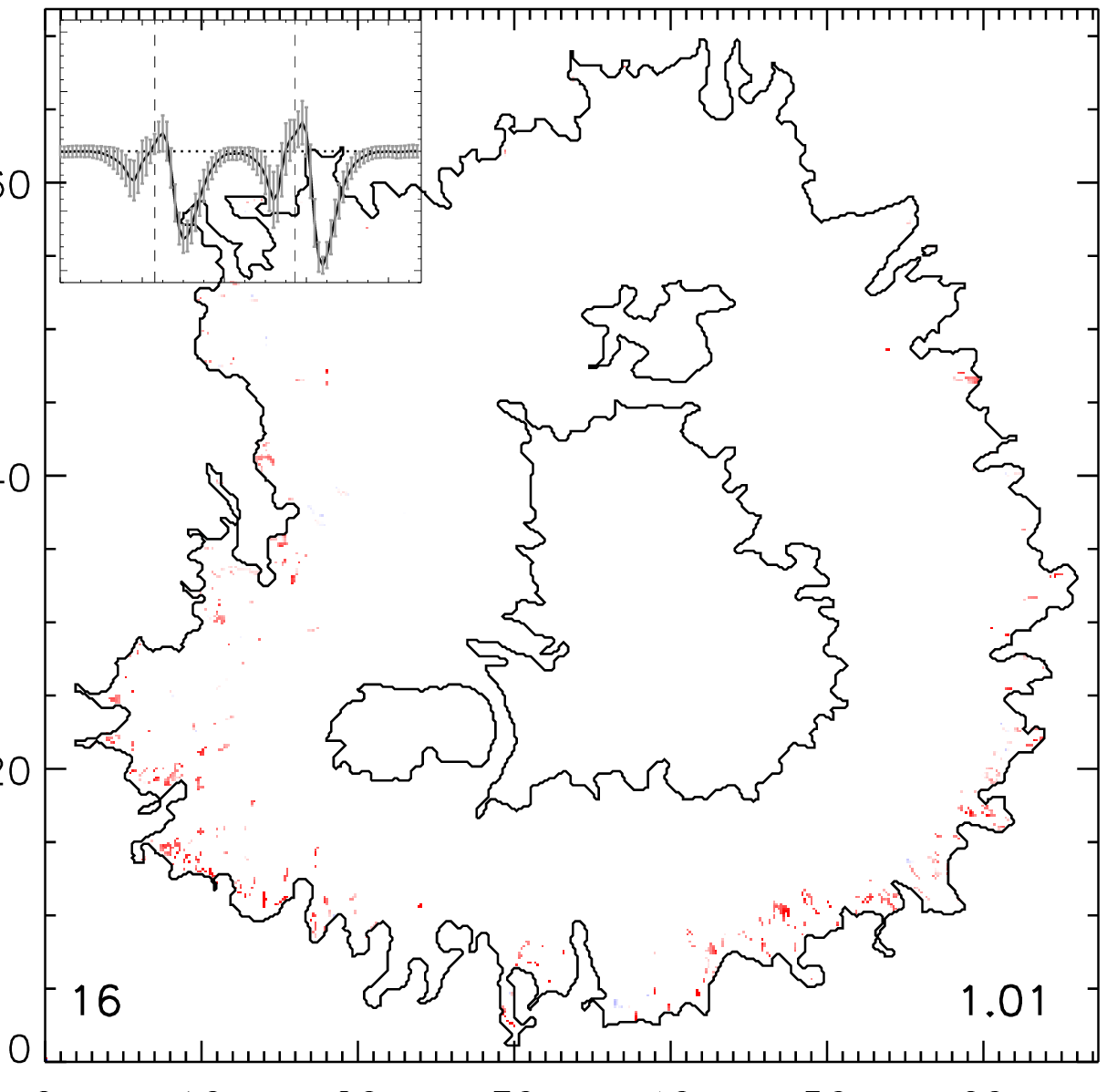}
\hspace{-70pt}
\includegraphics[angle=90,width = 0.45\textwidth]{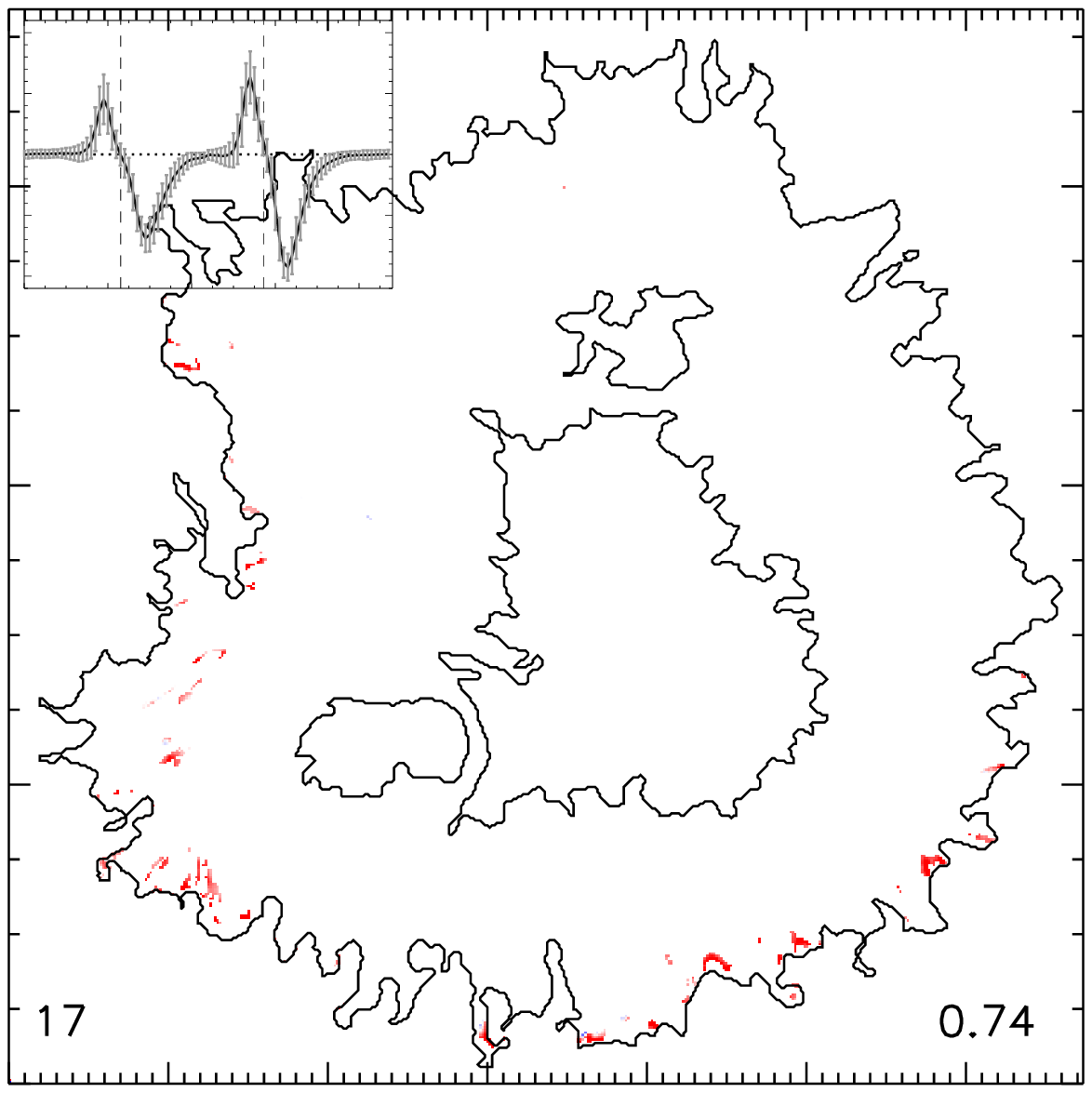}
\hspace{-70pt}
\includegraphics[angle=90,width = 0.45\textwidth]{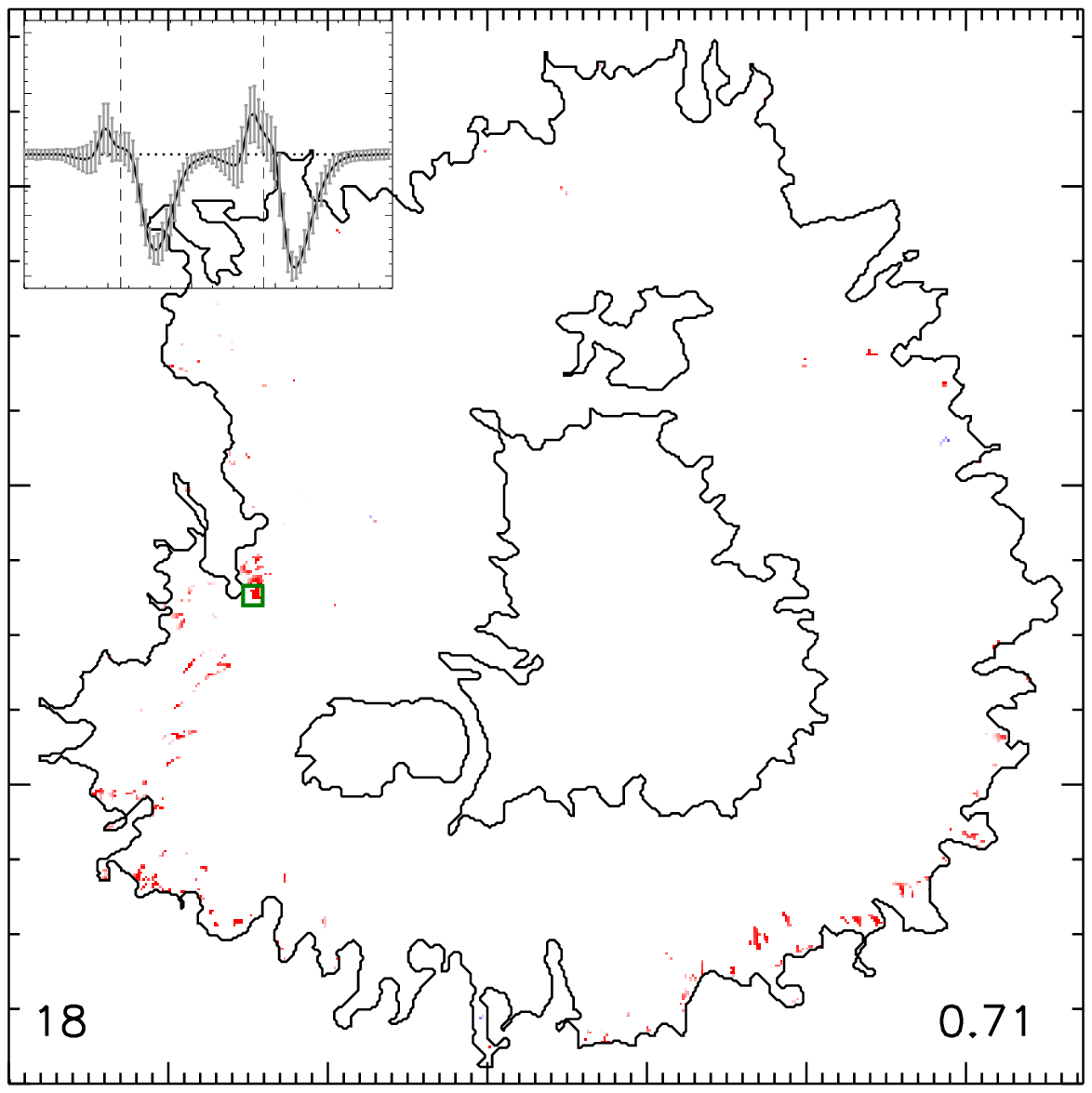}
}
\vspace{5pt}
\caption{Same as Fig.~\ref{fig04a} but for clusters 10 -- 18. The green square in 
Cluster 18 ($x=15$\arcsec, $y=33$\arcsec) indicates the location whose Stokes $V$ profile is
shown in Fig.~\ref{fig05}.}
\label{fig04b}
\end{figure*}

\begin{figure*}[!ht]
\centerline{
\hspace{50pt}
\includegraphics[angle=90,width = 0.45\textwidth]{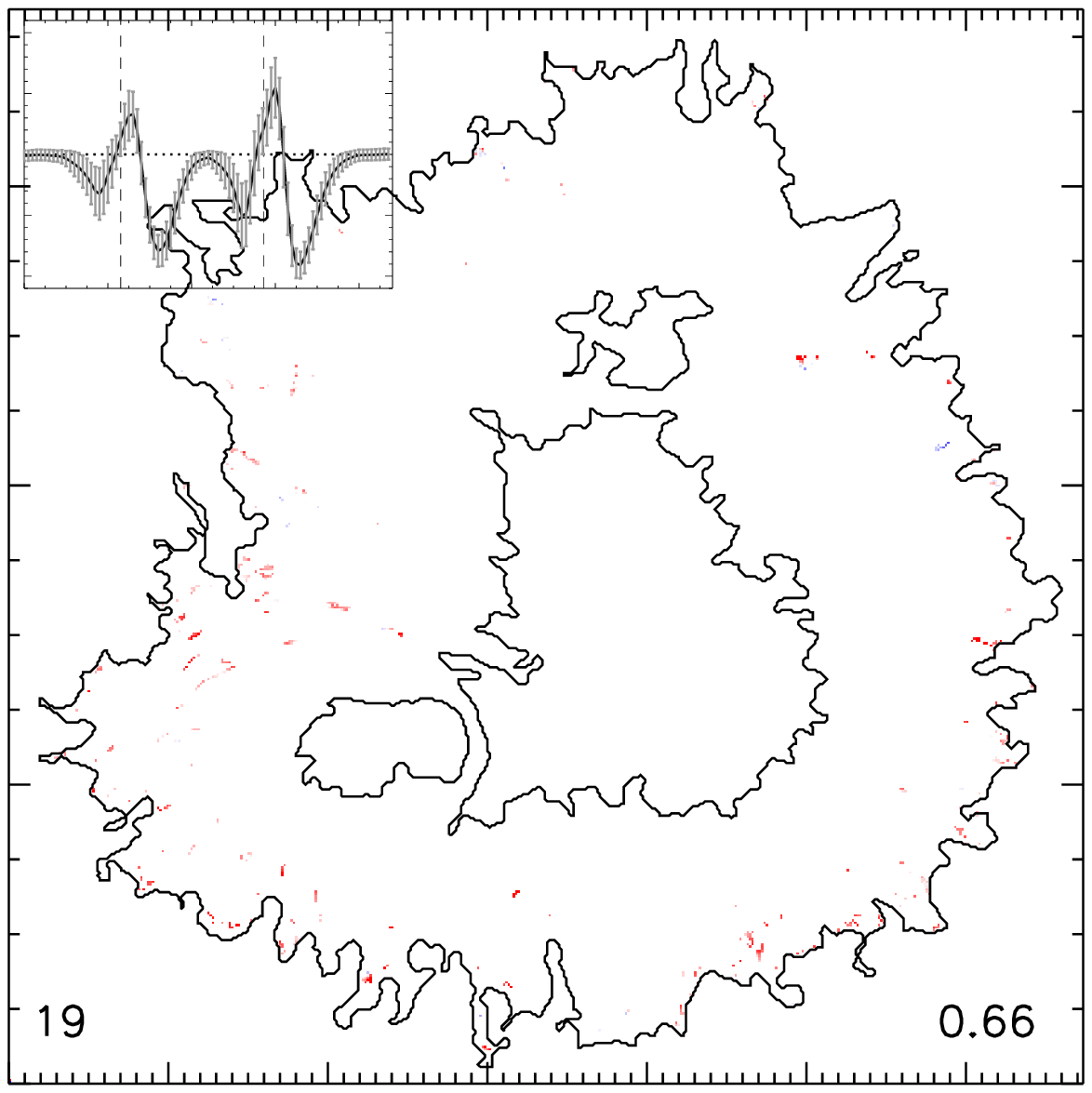}
\hspace{-70pt}
\includegraphics[angle=90,width = 0.45\textwidth]{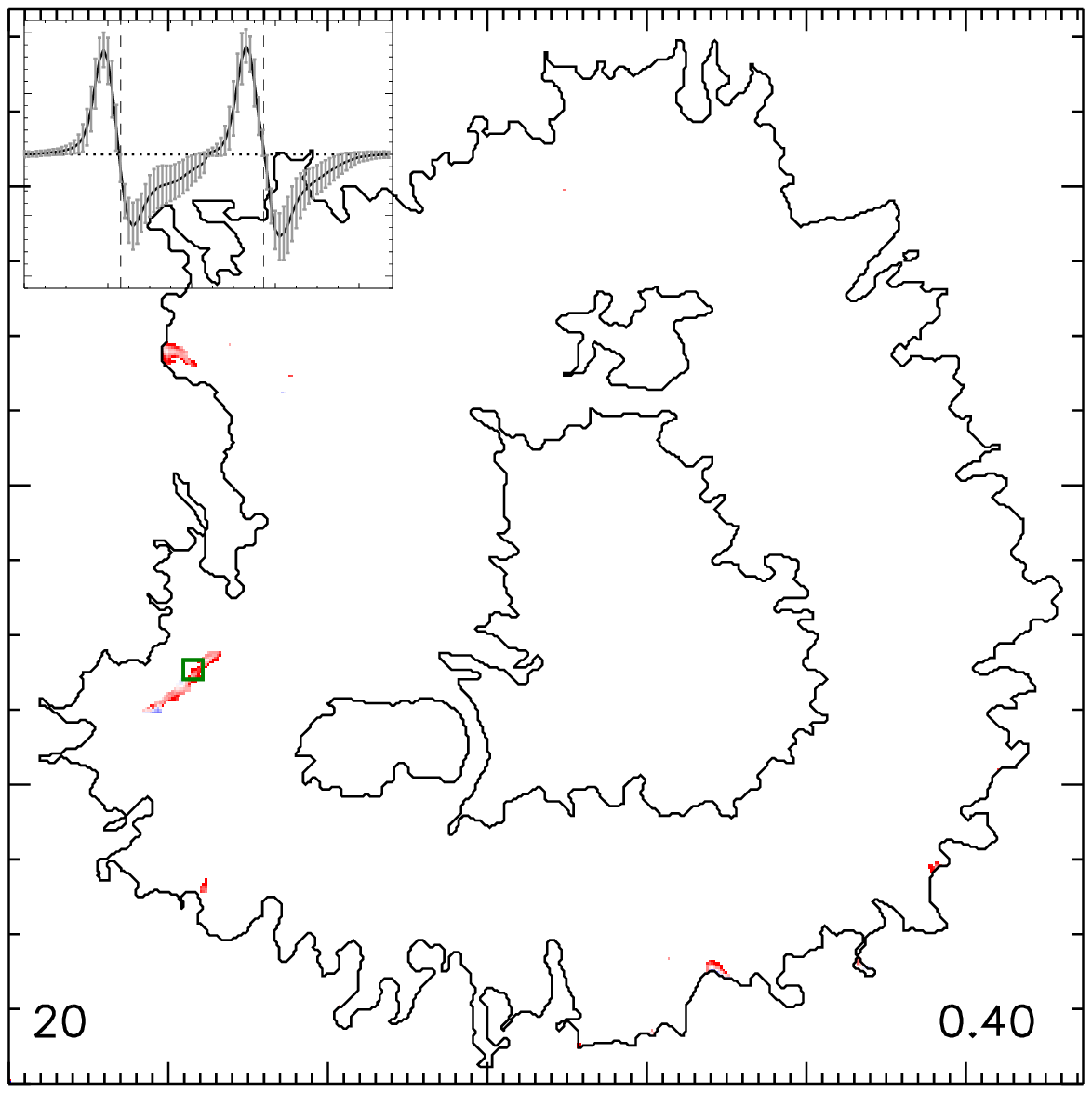}
\hspace{-70pt}
\includegraphics[angle=90,width = 0.45\textwidth]{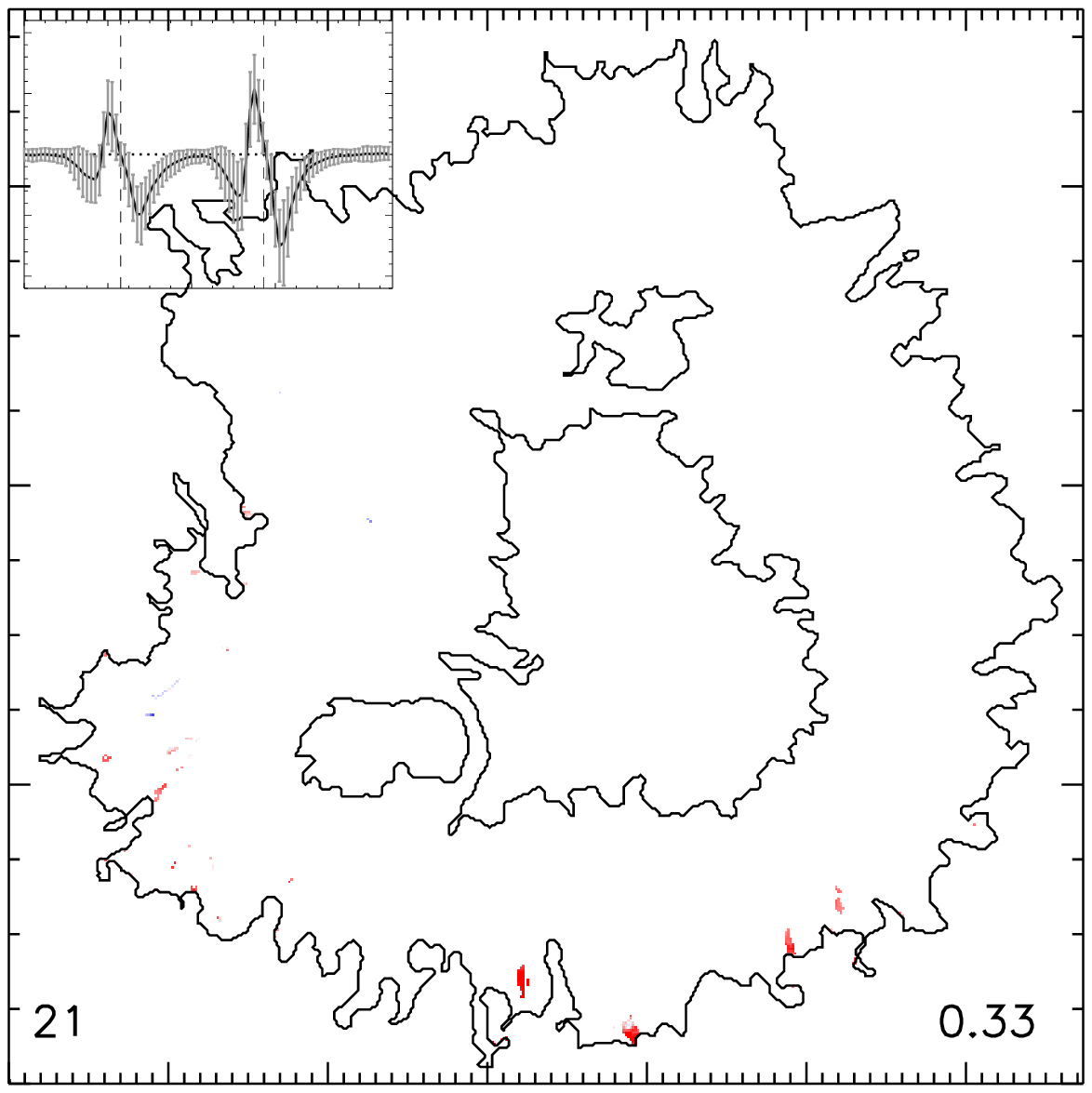}
}
\centerline{
\hspace{50pt}
\includegraphics[angle=90,width = 0.45\textwidth]{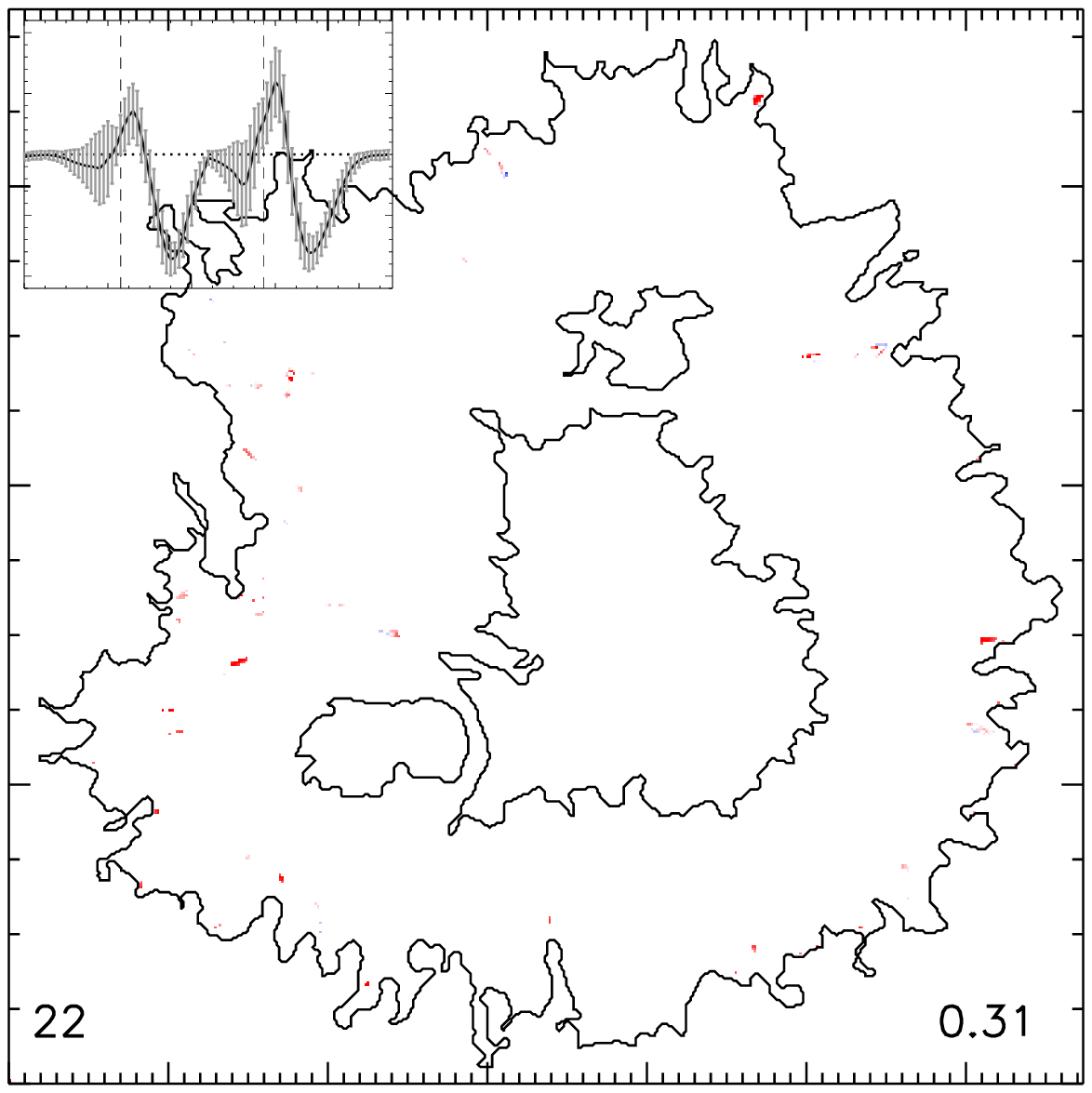}
\hspace{-70pt}
\includegraphics[angle=90,width = 0.45\textwidth]{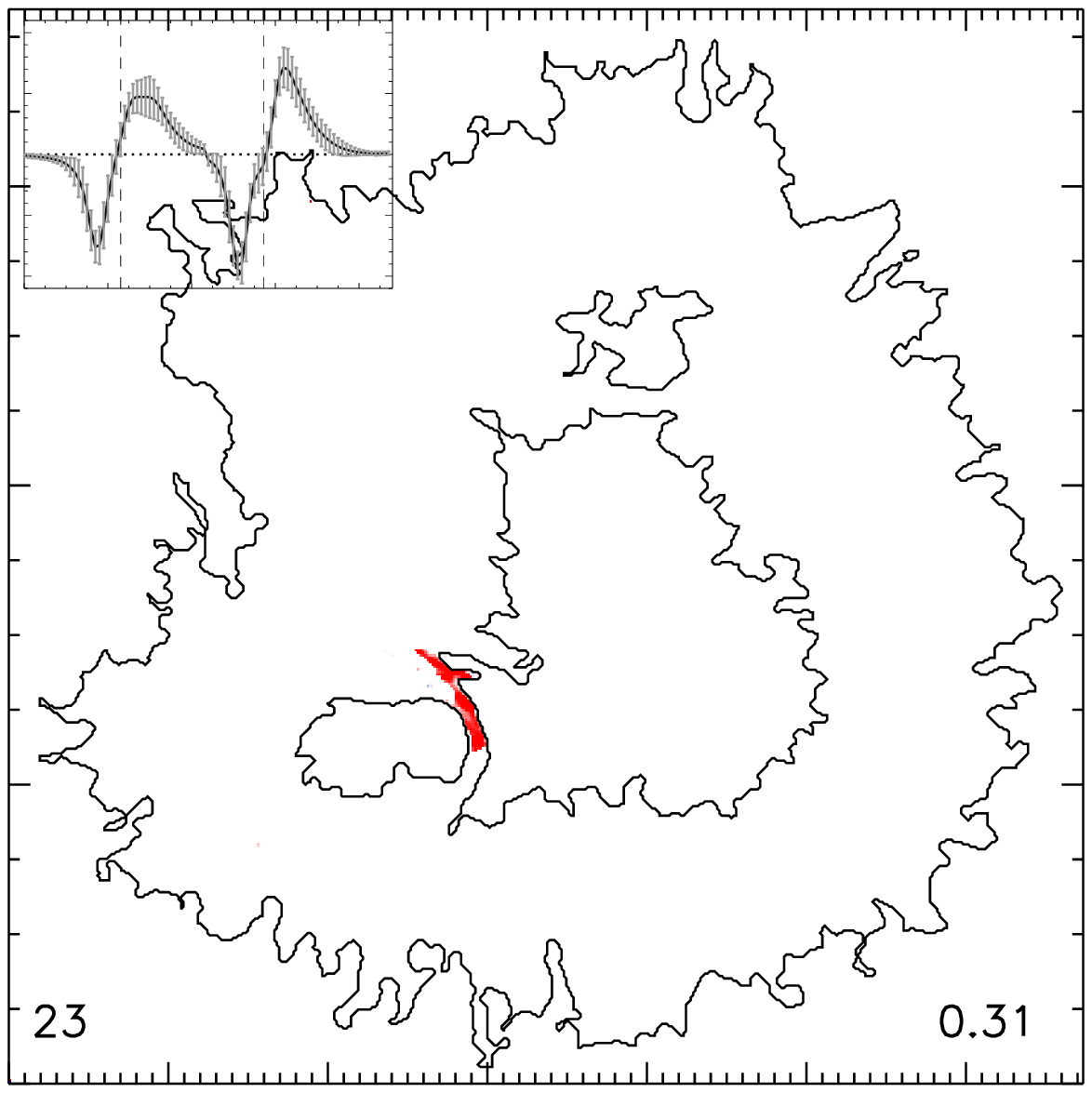}
\hspace{-70pt}
\includegraphics[angle=90,width = 0.45\textwidth]{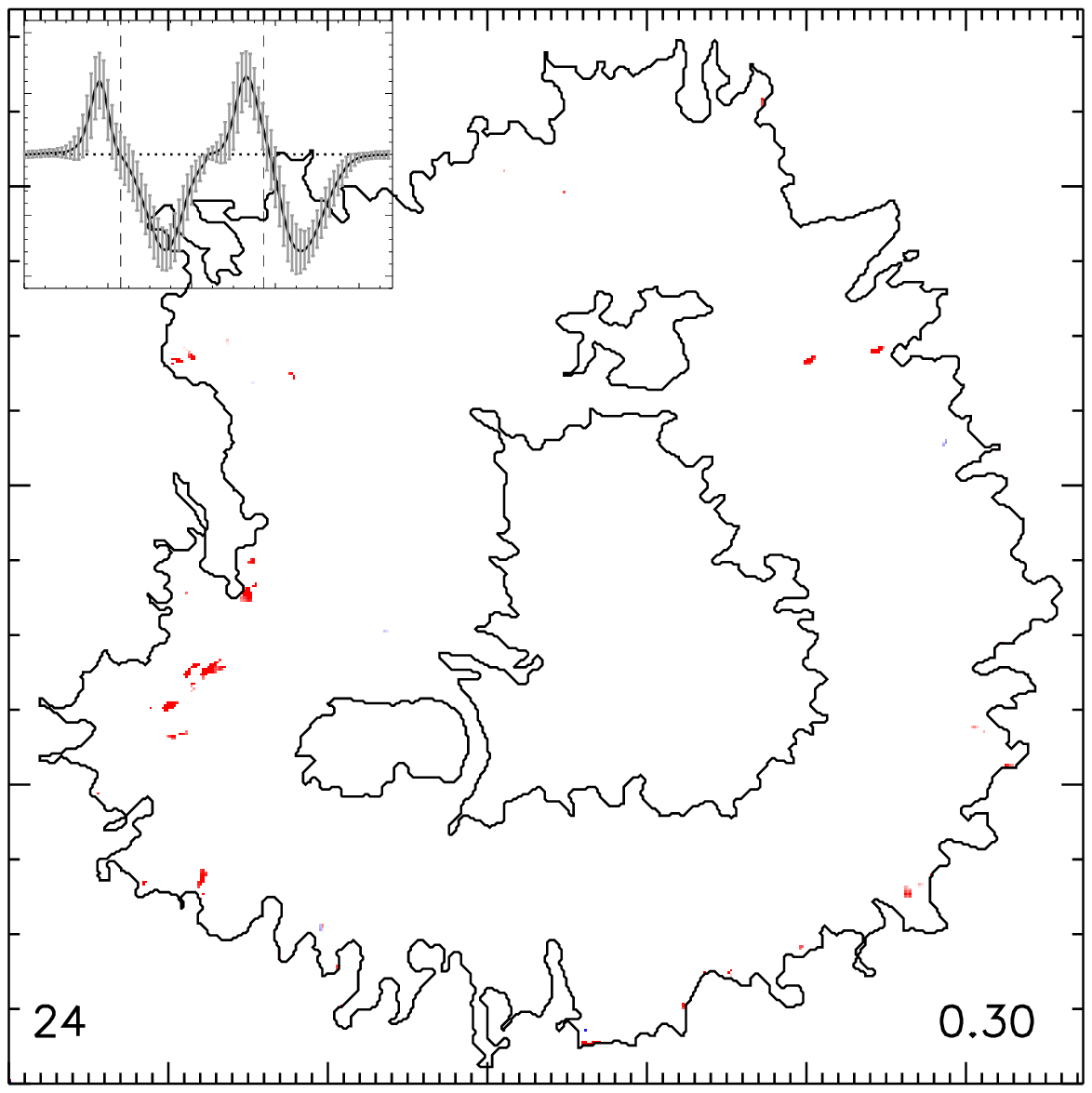}
}
\centerline{
\hspace{50pt}
\includegraphics[angle=90,width = 0.45\textwidth]{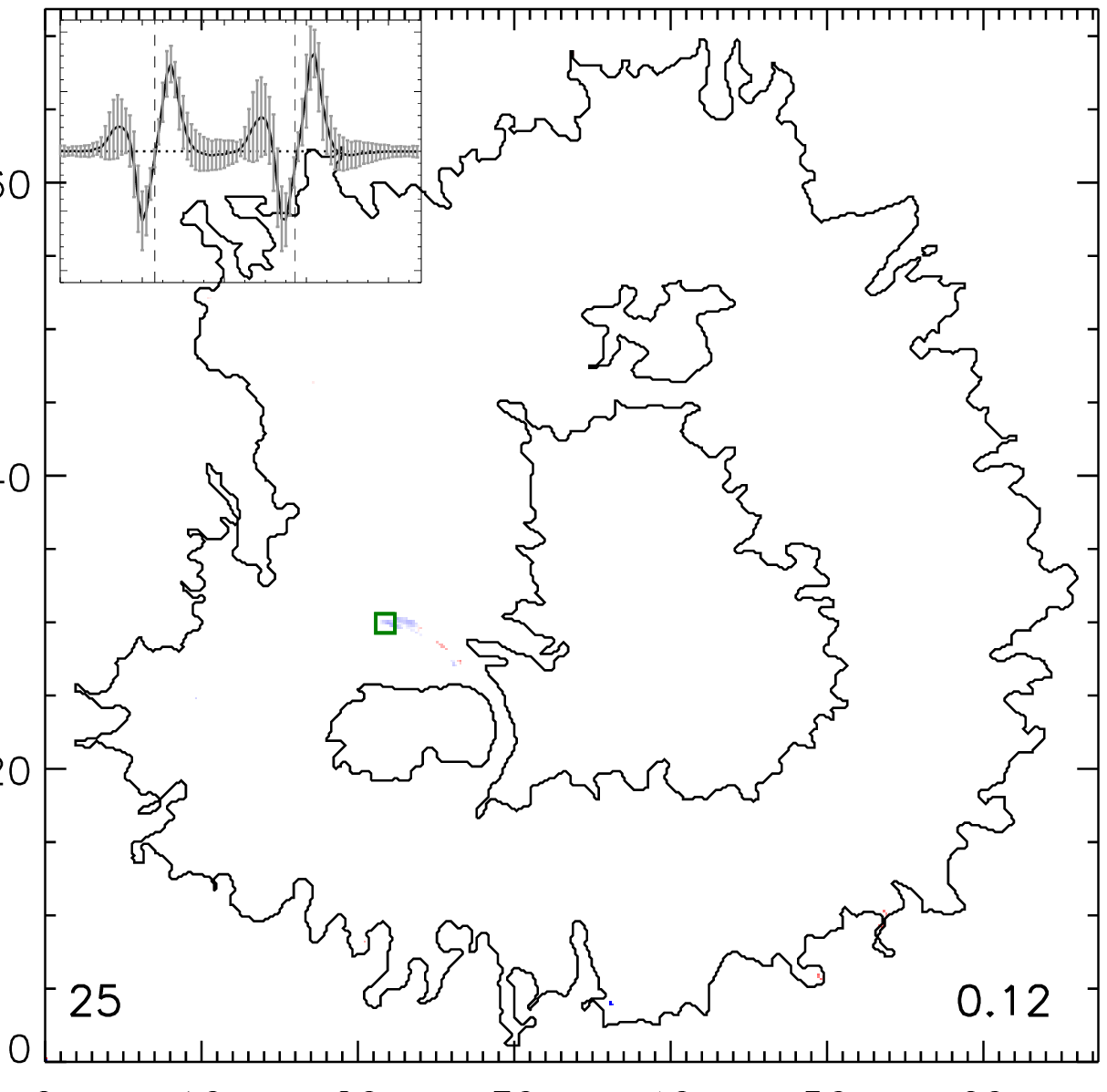}
}
\vspace{5pt}
\caption{Same as Fig.~\ref{fig04b} but for clusters 19 -- 25. The green squares in 
Cluster 20 ($x=11$\arcsec, $y=27$\arcsec), and Cluster 25 ($x=21$\arcsec, $y=30$\arcsec) indicate 
locations whose Stokes $V$ profiles are shown in Fig.~\ref{fig05}.}
\label{fig04c}
\end{figure*}

Once all the initial cluster centers were determined, the actual 
clustering process was carried out by computing the distance of each ($i^{\textrm\small{th}}$) Stokes $V$ profile from the 
$m^{\textrm\small{th}}$ cluster center $C$ as, 

\begin{equation}
D_{i,m} = \sqrt{\sum_{j=1}^{j=w} (C_{m,j} - V_{i,j})^2}
\end{equation}

\noindent{}and determining $m$ for which $D_{i,m}$ was a minimum. 
The index $j$ runs over the number of wavelength points, which in our case was $w = 88$, while
the index $m$ varies from 1 to $k$.  
The clustering algorithm was run for different number of clusters, ranging from 
$k=4$ to $k=36$. At the end of each clustering run, i.e. when the variance had converged, the clusters were 
re-ordered according to the number of profiles in each cluster and indexed from 1 to $k$, where 1 represents the 
most populous cluster, and  $k$ the most scarce. The total time for performing the clustering 
was about 14\,min on an iMac with a 3.1 GHz, 6-core, Intel i5 processor.
For each clustering run, where $k$ was varied, we computed the sum of 
the standard deviation of those clusters which contained nearly anti-symmetric Stokes $V$ profiles 
weighted by the number of profiles in that cluster. This was done on the basis that a significant part 
of the penumbra is known to contain profiles that do not exhibit very strong asymmetries.
The variation of the sum of the standard deviation with $k$ is shown 
in Fig.~\ref{fig02} where an exponential decaying function was used to determine the optimum choice of $k$
beyond which increasing the number of clusters did not yield any further change. 
The value of $k$ was chosen to be 25 using the elbow \citep{Thorndike1953} of the curve which works
best when the data set is clustered.

\section{Results}
\label{results}
\subsection{Inter-cluster Variations}
\label{inter}
Figure~\ref{fig03} shows the variation in cluster populations as the number of clusters change. For low cluster 
numbers, i.e., when {\tt{k}} $\le10$, the bulk of the penumbral area is occupied by the first couple of clusters and
the scarce clusters are confined to the outer regions of the penumbra. As {\tt{k}} increases, we find the above 
trend persists, with the populous clusters beginning to segregate into various sub-clusters. In addition, the filamentary
and small-scale structure, characteristic of the penumbra becomes more apparent when {\tt{k}} $\ge15$. The middle panel in 
the bottom row of the figure indicates that individual clusters can also be identified with specific features, such as the
southern filamentary LB with cluster no. 23, an elongated filament with cluster no. 20 ($x=11$\arcsec, $y=27$\arcsec), and the two, nearly 
circular patches in the north-western section of the penumbra with cluster no. 24 ($x=50$\arcsec, $y=48$\arcsec). 
The spatial distribution settles quite quickly between {\tt{k}} $=15$ and {\tt{k}} $=25$, which was
evident from Fig.~\ref{fig02}. Increasing the number of clusters beyond 25 only serves to divide the clusters further 
within themselves and does not produce any significant change in the spatial distribution.

\subsection{Spatial Distribution of Clusters for {\tt{k}} $=25$}
\label{spatial}
Figures~\ref{fig04a}--\ref{fig04c} shows the spatial distribution of individual clusters in the penumbra, when using {\tt{k}} $=25$,
with the mean profile for each cluster displayed in the top left corner. Since the clustering was performed on the profiles 
after removing the velocities, the binary map for each cluster was used to depict the corresponding LOS velocity as shown 
in the background. It is to be noted that the LOS velocity shown in the figures was derived from inversions 
that used height-independent parameters, hence the values are only representative of the average velocity in the height
forming region of the photospheric \ionn{Fe}{1} lines. Given the proximity of the sunspot to disc-centre we interpret the
blue- and red-shifts as upflows and downflows, respectively.

Clusters 1--5 and 7 (Fig.~\ref{fig04a}) cover nearly 74.3\% of the penumbra and comprise nearly, anti-aymmetric Stokes $V$ 
profiles. The variation of profiles among these clusters originate from the amplitude ratio between the two \ionn{Fe}{1} 
lines as well as the separation between the blue and red lobes. 
These profiles are associated with both upflows and downflows with mean values 
ranging from $-$0.1\,km\,s$^{-1}$ to $-$0.3\,km\,s$^{-1}$ and 0.1\,km\,s$^{-1}$ to 0.2\,km\,s$^{-1}$, respectively. The maximum 
values of upflows and downflows, on the other hand, range from $-$0.7\,km\,s$^{-1}$ to $-$1.4\,km\,s$^{-1}$ and 0.8\,km\,s$^{-1}$ 
to 1.3\,km\,s$^{-1}$, respectively. While upflows dominate in terms of area for clusters 1, 2, and 3, with values of 65\%, 
67\%, and 56\%, respectively, the downflowing regions exceed the upflowing areas by about 10\% for clusters 4, 5, and 7 
with values of 55\%, 54\%, and 57\%, respectively. 
Cluster 5, in particular, is solely confined to an annular region around the umbra-penumbra boundary, 
and mostly distributed over a line perpendicular to the sunspot line-of-symmetry (the line joining the spot center to the solar disc 
center). The average profile in cluster 5 is distinct from the other clusters in this group because the Stokes $V$ inversion 
at the zero crossing position in the \ionn{Fe}{1} line at 6302.5\,\AA\, is clearly visible, that arises from magneto-optical 
effects. The upflows and downflows in clusters 1 to 7 strongly bear the filamentary structure, the quintessential feature 
of the penumbra. 

\begin{table}[!h]
\caption{Summary of upflows and downflows in various clusters when {\tt{k}} $=25$. The area refers to the sunspot 
penumbra comprising 93458 pixels. The velocities are expressed in km\,s$^{-1}$.} \label{tab01}
\begin{center}
\resizebox{\columnwidth}{!}{
\begin{tabular}{r|rrr|rr|rr}
\hline\hline
No. & \multicolumn{3}{c|}{Area [\%]} & \multicolumn{2}{c|}{Upflows} & \multicolumn{2}{c}{Downflows} \\ \cline{2-8}
    & Total & Upflows & Downflows & Mean & Max. & Mean & Max \\
\hline\hline
 1 & 18.97 & 64.61 & 35.39 & -0.21 & -1.41 &  0.12 &  0.87 \\
 2 & 15.26 & 67.20 & 32.80 & -0.23 & -1.00 &  0.20 &  1.07 \\
 3 & 14.55 & 56.08 & 43.92 & -0.16 & -1.00 &  0.14 &  1.27 \\
 4 & 11.84 & 45.08 & 54.92 & -0.29 & -1.17 &  0.22 &  1.09 \\
 5 &  9.85 & 45.52 & 54.48 & -0.10 & -0.67 &  0.09 &  0.82 \\
 6 &  8.27 & 32.88 & 67.12 & -0.25 & -1.16 &  0.30 &  1.15 \\
 7 &  3.79 & 42.55 & 57.45 & -0.18 & -1.04 &  0.18 &  1.31 \\
 8 &  2.60 & 18.12 & 81.88 & -0.20 & -0.81 &  0.36 &  1.20 \\
 9 &  1.95 & 38.14 & 61.86 & -0.19 & -1.14 &  0.29 &  0.98 \\
10 &  1.95 & 33.21 & 66.79 & -0.27 & -1.56 &  0.30 &  1.04 \\
11 &  1.58 & 67.93 & 32.07 & -0.31 & -1.69 &  0.17 &  0.81 \\
12 &  1.26 &  5.08 & 94.92 & -0.15 & -0.87 &  0.59 &  2.36 \\
13 &  1.17 & 13.87 & 86.13 & -0.15 & -1.07 &  0.39 &  1.17 \\
14 &  1.04 & 23.55 & 76.45 & -0.17 & -0.97 &  0.34 &  1.09 \\
15 &  1.01 & 14.87 & 85.13 & -0.20 & -0.71 &  0.39 &  1.39 \\
16 &  1.01 &  5.28 & 94.72 & -0.09 & -0.26 &  0.50 &  1.65 \\
17 &  0.74 &  3.76 & 96.24 & -0.12 & -0.55 &  0.73 &  2.35 \\
18 &  0.71 &  3.30 & 96.70 & -0.16 & -0.59 &  0.73 &  3.23 \\
19 &  0.66 & 10.97 & 89.03 & -0.15 & -0.69 &  0.48 &  2.65 \\
20 &  0.40 & 11.38 & 88.62 & -0.21 & -0.57 &  0.68 &  2.89 \\
21 &  0.33 & 10.26 & 89.74 & -0.17 & -0.72 &  0.57 &  1.28 \\
22 &  0.31 & 14.97 & 85.03 & -0.17 & -0.75 &  0.59 &  4.46 \\
23 &  0.31 &  3.41 & 96.59 & -0.09 & -0.30 &  1.10 &  3.75 \\
24 &  0.30 &  5.63 & 94.37 & -0.24 & -0.89 &  1.47 &  5.94 \\
25 &  0.12 & 69.44 & 30.56 & -0.19 & -1.20 &  0.36 &  0.79 \\
\hline \hline
\end{tabular}
}
\end{center}
\end{table}

Cluster 11 (Fig.~\ref{fig04b}) consists 
of profiles with an extended blue lobe and is present in 1.6\% 
of the penumbral area. As a result, this cluster comprises 68\% upflows with a maximum value of $-$1.7\,km\,s$^{-1}$.

The next family of profiles consist of three, clear lobes, resembling Stokes $Q$ with the third lobe having an opposite sign 
as the spot and on the red-side of the profile. This spectral feature is seen in clusters 8 to 19 
(Figs.~\ref{fig04b} and \ref{fig04c}), excluding clusters 11, 12, 17, and 18 and 
account for 11\% of the penumbral area. The velocities in these 8 clusters are predominantly downflows with upflows
seen in clusters 9 and 10 and occupying 38\% and 33\% of the patch area. In addition, the downflows are confined to smaller 
patches or blobs while the filamentary structure is only apparent in cluster 8. The downflows in this family of clusters have 
maximum values exceeding 1\,km\,s$^{-1}$ and are as large as 2.7\,km\,s$^{-1}$ in cluster 19. 
Cluster 6 can also be regarded as a part of this family of profiles, as the mean profile exhibits a very weak, although 
discernible, third lobe of opposite polarity and is mostly confined to the outer penumbra. This cluster has an area of 
about 8\% in the penumbra with downflows occupying 67\% of it while the maximum upflows and downflows are about 1.2\,km\,s$^{-1}$.
To this family of three lobed profiles, one can include clusters 21 and 22 (bottom row of the figure) with the exception that 
there is a large scatter associated with profiles occupying an area of 0.33\% and 0.31\% of the penumbral area.
Cluster number 22, in particular has the second highest downflows reaching values of about 4.5\,km\,s$^{-1}$.

The remaining clusters account for about 4\% of the penumbral area and consist of four prominent groups/families.  
Clusters 12 and 18 comprise predominantly of a red lobe, occupying 2\% of the penumbral area and are seen mostly at the limb-ward 
side of the sunspot's boundary. The downflows in the two clusters reach a maximum value of 2.4\,km\,s$^{-1}$ and 3.2\,km\,s$^{-1}$, 
respectively. 

Cluster 23 (Fig.~\ref{fig04c}) is seen exclusively in the filamentary LB in the southern part of the sunspot, occupying an area 
of 0.3\% of the penumbra with an extended red lobe but with the same sign as the spot. The downflows in the LB are about 
3.8\,km\,s$^{-1}$.

Clusters 17 (Fig.~\ref{fig04b}) and 24 (Fig.~\ref{fig04c}) exhibit highly asymmetric profiles with two primary 
lobes but with a sign opposite that of the spot. These clusters 
appear in specific patches as downflows with values of 2.4\,km\,s$^{-1}$ and 5.9\,km\,s$^{-1}$, respectively and occupy 1\% of 
the penumbral area. Cluster 20 can be considered a sub-category of clusters 17 and 24, as it shares the same sign of the lobes 
as the others but consists of a highly extended red-lobe. One of the locations of this cluster consists of a fairly conspicuous 
filamentary structure ($x=11$\arcsec, $y=27$\arcsec) that is also spatially associated with the other two clusters in this family. 

The final remaining cluster is number 25 (Fig.~\ref{fig04c}) that occupies about 
0.12\% of the penumbral area. The profiles in this cluster have an opposite sign to those seen in the second family of profiles 
(8--19) consisting of three lobes and associated with strong downflows. This cluster mostly exhibits weak upflows located close 
to the entrance of the southern filamentary LB. The above values are summarized in Table~\ref{tab01}.

\subsection{Stokes $V$ Spectra Associated with Select Clusters}
\label{spectra}
\begin{figure}[!h]
\centerline{
\includegraphics[angle=0,width = \columnwidth]{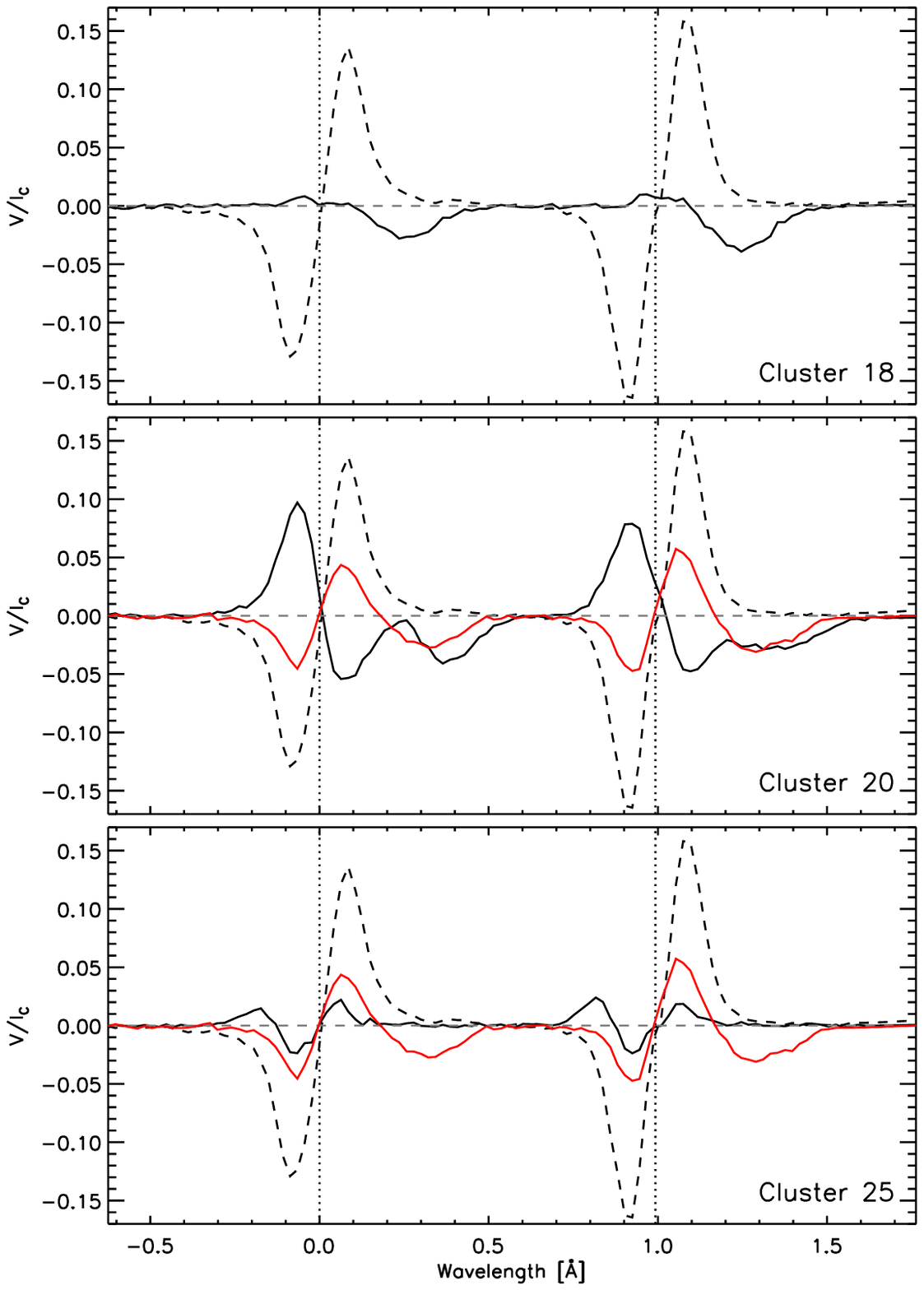}
}
\vspace{-15pt}
\caption{Stokes $V$ profiles in different clusters (solid black line) versus a normal anti-symmetric penumbral profile (dashed black 
line). The red line corresponds to the Evershed flow returning to the photosphere, whose location is indicated by the black circle in 
Fig.~\ref{fig01}. The gray horizontal dashed line corresponds to an amplitude of zero.}
\label{fig05}
\end{figure}

We now look at the actual spectra from select clusters and compare them with the normal, anti-symmetric $V$ profile that dominates
the penumbra. As mentioned earlier, this is to ascertain their amplitude since the clustering was performed after the profiles were 
normalized to their individual unsigned maxima. Figure~\ref{fig05} shows Stokes $V$ profiles from clusters 18, 20, and 25, with 
their location marked in Figs.~\ref{fig04b} and \ref{fig04c} with a green square. The top panel shows that the Stokes $V$ profile 
in cluster 18 is primarily dominated by a single red lobe with an amplitude of about 4\% and has a sign opposite to that of the 
sunspot in comparison to the stronger, anti-symmetric Stokes $V$ profile (dashed line). 

The middle panel shows the $V$ profile in the large elongated patch belonging to cluster 20, where the presence of 2 separate
red lobes indicates strong, possibly supersonic downflows. The difference between these profiles versus those typically 
associated with the Evershed flow returning to the photosphere (red line) is that the slower or rest component also has a polarity opposite 
that of the sunspot as seen from the sign of the blue lobe. The strength of the strongly red-shifted lobes is about 4\% and 3\% 
in the two \ionn{Fe}{1} lines, respectively.

Finally, the bottom panel of the figure shows the profile associated with a patch exhibiting weak upflows seen in cluster 25 with an 
amplitude of about 2\% and comprises three lobes. The rest component has the same polarity as the sunspot, but the third lobe associated
with the weak upflow is of opposite polarity. Thus, these profiles are of opposite sign to that seen in the Evershed flow returning 
to the photosphere (red line), the velocities are mostly upflows, and considerably smaller in magnitude.

\section{Summary and Discussion}
\label{discuss}
The segregation of Stokes $V$ profiles using the {\tt{k-means}} clustering method reveals that nearly three-quarters
of the profiles in a penumbra of a regular, unipolar sunspot are nearly normal and anti-symmetric with two, well defined lobes. 
These profiles comprise both upflows and downflows with mean values of about $-$0.3\,km\,s$^{-1}$ and 0.2\,km\,s$^{-1}$, 
respectively. This set of profiles can be found in a total of six clusters when using {\tt{k}} $ = 25$ for the clustering.
In comparison, such normal, anti-symmetric profiles only constitute about 7\% of the area in the QS \citep{2011A&A...530A..14V} 
while having an area five times larger than the penumbra. Thus, employing a ME approximation for the inversion of these profiles
can be straightforward and simple. 

The second dominant class of profiles in the penumbra comprise three lobes, similar to Stokes $Q$ accounting for about 21\% 
of the penumbral area and comprising 11 clusters. In the QS, such profiles occupy about 8\% of the area \citep{2011A&A...530A..14V}. 
These profiles dominantly show downflows which are associated with the Evershed flow returning to the photosphere, 
\citep{1997Natur.389...47W,2001ApJ...549L.139D,2004A&A...427..319B,2008A&A...480..825B,2003A&A...410..695M,2007PASJ...59S.593I,
2013A&A...549L...4R}, wherein the rest component has the same polarity as the spot and the high-speed, downflowing component 
has the opposite polarity. Cluster 6 has a similar property to the above, although the third lobe is extremely weak, but 
nevertheless discernible.

The remaining penumbral region of about 4\% is dominated by four categories of profiles. The first are the strong downflows, 
sometimes reaching supersonic values, seen entirely in the filamentary LB \citep{2009ApJ...704L..29L,2011ApJ...727...49L} 
and in a single cluster. The polarity of both components is the same as the parent sunspot. However, we do not see similar 
profiles in the northern granular LB which would suggest that these anomalous profiles are not a consequence of overturning 
convection in the weakly magnetized LB. Moreover, the magnetic field strength of the rest and supersonic component 
is about 2.0\,kG and 1.7\,kG, respectively and nearly vertical in both cases. \citet{2009ApJ...704L..29L} had shown that if the
Stokes profiles in the LB were fitted with discontinuities in the physical parameters, the high-speed downflows would occur in the
deeper layers below $\log\tau=-0.5$. The high speed downflows could result from the gravitational acceleration of gas along 
vertical flux tubes or due to the iso-$\tau$ level associated with the Wilson depression sampling deeper layers where the 
plasma velocity would be higher. In addition, these supersonic downflows in LBs are known to be associated with 
long-lived jets/arc-like brightenings in the chromosphere, although the causal relation between the two is still an open 
question \citep{2008SoPh..252...43L}.

The second category comprises profiles with a dominant or single red lobe that cover about 1.4\% of the penumbral area. Such 
profiles are also seen in the QS and account for about 11\% of the area \citep{2011A&A...530A..14V}. 
These single-lobed profiles are known to be associated with emerging $\Omega$ loops \citep{2012ApJ...747L..36V}
in the QS, with earlier studies focusing on the topology, dynamics 
\citep{2007ApJ...666L.137C,2007A&A...469L..39M,2009ApJ...700.1391M,2010ApJ...714L..94M}, and specific examples of 
emergence \citep{2010ApJ...713.1310I}. The emergence of magnetic flux in the QS 
\citep{2009ApJ...700.1391M,2014ApJ...781..126O,2016ApJ...825...93O,2021ApJ...911...41G} 
or in sunspots \citep{2015A&A...584A...1L,2021A&A...652L...4L} plays a vital role in heating the overlying chromosphere 
depending on the pre-existing field configuration. 
These single-lobed 
profiles can also be associated with the submergence of magnetic flux in the QS as shown in \citet{2012ApJ...748...38S} 
using inversions and numerical simulations, where the fraction of such profiles in the QS was about 2\%.

An inversion of these Stokes profiles was performed using height-dependent 
parameters, using three nodes for the magnetic field strength, inclination, and LOS velocity, while two nodes were provided 
for the azimuth. One finds that the field inclination 
is about $50^\circ$ at $\log\tau=0$ and changes to nearly $90^\circ$ at $\log\tau=-1.5$, while the LOS velocity is at rest at $\log\tau=-1.5$
and increases to about 7\,km\,s$^{-1}$ at $\log\tau=0$. While this configuration is consistent with one leg of the $\Omega$ loop hosting 
downflows, a visual inspection of the profiles in the vicinity of the single, red-lobed profiles did 
not reveal any single, blue-lobed profile, nor was there any indication of the latter when using higher values of {\tt{k}} $=36$.
We suggest that instead of smooth gradients, employing discontinuities along the LOS would be a better alternative to reproduce
such single-lobed profiles.

The third category of profiles comprises three lobes, however, both the rest and strongly red-shifted component have a polarity opposite
that of the sunspot, unlike those associated with the Evershed flow returning to the photosphere. A two-component inversion of the Stokes
profiles using height-independent parameters indicates that the component hosting supersonic downflows has a fill fraction of about 
26\%, while both the rest and high-speed components have similar field strengths and inclinations of about 1500\,G and $40^{\circ}$,
(with respect to the vertical) respectively. However, a more likely scenario is that these components are stacked in height 
with the high-speed downflows located in the deeper layers. One possible explanation as to why the rest component in these types
of profiles is also of opposite polarity as the sunspot could be the line-of-sight sampling a magnetized atmosphere with the field 
oriented in the opposite direction throughout as opposed to the returning EF where the neighboring, nearly vertical, 
intra-spine fields arch over the high-speed EF channels. This is supported by the spatial distribution and morphology of the pixels
which harbor these profiles which are predominantly 
located in large, isolated and extended structures (cluster 20 in Fig.~\ref{fig04c}) while the EF downflows are seen in 
diffuse/smaller, but nearly ubiquitous patches or blobs in the outer limb-side penumbra (e.g. clusters 13--15 in Fig.~\ref{fig04b}).

The final category of profiles, while comprising three lobes, has the opposite sign as profiles exhibiting the Evershed flow 
returning to the photosphere. A two-component inversion similar to the category above show that the dominant component, having 
a fill fraction of about 86\%, has the same polarity as the spot and is at rest, while the second component consists of upflows 
of about $-$1.5\,km\,s$^{-1}$ and has an opposite polarity as the sunspot. While both components are of opposite polarity, 
they are nevertheless highly inclined which explains the reduced amplitude of the Stokes $V$ signal. As these profiles are 
located close to the entrance of the filamentary LB, they could be magnetically connected to the large active region 
filament, seen in H$\alpha$, located outside the sunspot and extending towards the LB 
\citep{2008ApJ...673L.215O,2009ApJ...697..913O,2011ApJ...738...83S}. Intermittent, strong chromospheric brightenings 
were observed in localized patches on the filament channel as well as over the LB \citep{2009ApJ...697..913O}. 
A similar scenario was found by \citet{2013ApJ...770...74K}, where the 3-lobe Stokes \textit{V} 
profiles were also identified in an umbral filament, which was also connected to magnetic structures outside of the 
sunspot through the overlying chromosphere. Those 3-lobe Stokes \textit{V} profiles were also inverted considering two 
magnetic components. 

Some, if not all, of the anomalous Stokes profiles described here are the photospheric manifestations of either dynamics or 
structures in the overlying chromosphere, although the possible causal relation between the two may be difficult to ascertain when 
the information on the magnetic fields and plasma motions is only available in the photosphere. Further investigations are 
needed to determine the physical mechanisms producing these small-scale inhomogeneities by combining observations with radiative
magneto-hydrodynamic simulations. However, the speed and simplicity 
of {\tt{k-means}} makes it a viable tool to 
identify such anomalous profiles and also track them in time, not only in the photosphere but in higher atmospheric layers as well. 
At present, {\tt{k-means}} is widely employed in the IRIS Inversion based on Representative profiles Inverted by 
STiC \citep[IRIS$^{\textrm{\tiny{2}}}$;][]{2019ApJ...875L..18S}. 
A similar strategy could be implemented in the CAlcium Inversion based on a Spectral ARchive 
\citep[CAISAR;][]{2015ApJ...798..100B,2019ApJ...878...60B}

The clusters derived in this study are specific for a sunspot located close to disc center. As the EF can be perceived as hot 
upflows in the inner penumbra that turn into a nearly horizontal outflow and are embedded in a nearly vertical, static magnetic field, 
the LOS will sample different parts of this magnetic geometry as the heliocentric angle increases \citep{2002A&A...381..668S}. 
Thus, one would expect the number of profiles with strong asymmetries to increase as the viewing angle changes 
from disc center to the limb. Nevertheless, with {\tt{k-means}} one can ascertain the nature and spatial distribution of the clusters 
as a function of heliocentric angle for spots that have a similar configuration as the one analyzed in this study.

We finally note that the clustering was carried out on the Stokes spectra without a correction for the spatially coupled polarized 
stray light due to Hinode SOT's point spread function \citep{2012A&A...548A...5V}. The above correction would render the weaker 
spectral anomalies more prominent that would alter the mean profile in the clusters as well as possibly introduce additional 
cluster families to the ones described above, but the minority clusters identified above would remain unaffected.

\section{Conclusions}
\label{conclude}
The usage of {\tt{k-means}} as a tool to classify Stokes $V$ spectra in a regular sunspot is a quick and easy approach to identify 
different families of profiles and their spatial occurrences. In the regular sunspot, 75\% and 21\% of the penumbral area is occupied 
by normal, anti-symmetric profiles with two lobes and three-lobed profiles, respectively. The latter in particular is the well-known 
profile associated with the Evershed flow returning to the solar photosphere where the strong, downflowing component often has a 
polarity opposite to the sunspot. The remaining penumbral area of 4\% is dominated by four groups/families wherein one 
of them comprises Stokes $V$ profiles with only one red lobe while the other three comprise three lobes with different polarity 
combinations of the two magnetic components present in the profile, making them distinct from those associated with the Evershed 
flow returning to the photosphere. Further investigations are needed to understand the physical mechanisms that produce such 
spectra in sunspots and their possible coupling to the overlying chromosphere.

\section*{Acknowledgments}
Hinode is a Japanese mission developed and launched by ISAS/JAXA, collaborating with NAOJ as a domestic partner, NASA, and 
STFC (UK) as international partners. Scientific operation of the Hinode mission is conducted by the Hinode science team 
organized at ISAS/JAXA. This team mainly consists of scientists from institutes in the partner countries. Support for 
the post-launch operation is provided by JAXA and NAOJ (Japan), STFC (UK), NASA, ESA, and NSC (Norway). 
The SIR inversions were performed using the 100TF cluster Vikram-100 at Physical Research Laboratory, India.
We thank the anonymous referees for reviewing our article and providing useful comments.


\bibliographystyle{jasr-model5-names}
\biboptions{authoryear}

\bibliography{louis-ref}

\end{document}